\documentclass[preprint,12pt]{elsarticle}
\usepackage{amssymb}
\usepackage{graphicx}
\journal{Nuclear Physics A}

\begin{document}

\begin{frontmatter}

\title{Physical properties of  Polyakov loop geometrical clusters in  SU(2) gluodynamics}

\author[r]{A.I.~Ivanytskyi}
\author[r]{K.A.~Bugaev}
\author[p]{E. G.~Nikonov} 
\author[q]{E.-M.~Ilgenfritz}
 \author[r,els]{D.R. Oliinychenko}
 \author[r,t]{V.V.~Sagun}
\author[els,foc]{I.N.~Mishustin} 
\author[r]{V.K.~Petrov}
\author[r]{G.M.~Zinovjev}

\address[r]{Bogolyubov Institute for Theoretical Physics,
National Academy of Sciences of Ukraine,Metrologichna str. 14$^b$, Kiev-03680, Ukraine}
\address[p]{Laboratory for Information Technologies, JINR, 141980 Dubna, Russia}
\address[q]{Bogoliubov Laboratory of Theoretical Physics, JINR, 141980 Dubna, Russia}
\address[els]{Frankfurt Institute for Advanced Studies (FIAS), Goethe-University, Ruth-Moufang Str. 1, 60438
Frankfurt upon Main, Germany}
\address[t]{CENTRA, Instituto Superior T$\acute{e}$cnico, Universidade de Lisboa,
Av. Rovisco Pais 1, 1049-001 Lisboa, Portugal}
\address[foc]{Kurchatov Institute, Russian Research Center, Akademika Kurchatova Sqr., Moscow, 123182, Russia}

\begin{abstract}
We apply the liquid droplet model to describe  the clustering phenomenon
in SU(2) gluodynamics, especially, in the vicinity of the deconfinement phase transition.
In particular, we analyze the size distributions of clusters formed by the Polyakov loops of the same 
sign. Within such an approach this  phase transition  can 
be considered as the transition between two types of liquids where
one of the liquids (the largest droplet of a certain Polyakov loop sign) 
experiences a condensation, while the other one   (the next to largest droplet 
of opposite Polyakov loop sign) evaporates. The clusters of smaller sizes 
form two accompanying gases, and their 
size distributions  are described by the liquid droplet parameterization.
By fitting the lattice data we have extracted 
the  value of Fisher exponent $\tau =$ 1.806 $\pm$ 0.008. 
Also we found that  the temperature dependences of the 
 surface tension of both gaseous clusters  are entirely different  below and above  the phase transition 
and, hence, they  can serve as an order parameter. 
The critical 
exponents of the surface tension coefficient in the 
vicinity of the phase transition are found. 
Our analysis  shows  that the  temperature dependence of
the  surface tension coefficient above the critical temperature 
has a $T^2$ behavior 
 in one gas of clusters and  $T^4$ in the other one. 
\end{abstract}

\begin{keyword}

Key words: Polyakov loops, clusters, liquid droplet formula,  surface tension, new order parameters \\
PACS:25.75.Nq, 25.75.-q 

\end{keyword}

\end{frontmatter}

\section{Introduction}

The lattice formulation of quantum chromodynamics (QCD) is the only first 
principle tool which allows us to study  the phase 
transformations between the confined and deconfined state of strongly interacting 
matter. Despite great successes of lattice QCD, up to now there is no complete 
understanding of the deconfinement phase transitions (PT) or crossover 
phenomena. A rich collection of results was obtained using the 
guidance of the Svetitsky-Jaffe hypothesis \cite{Svetit:I, Svetit:II}.  
This hypothesis links the deconfinement PT of a SU(N) pure gauge theory
in (d+1)--dimensions to the magnetic transition in the corresponding 
d--dimensional symmetric spin system being invariant under global center group 
Z(N)-transformations, actually in two versions: with SU(N) valued spins or with 
Z(N) valued spins.
The role of spins \cite{Polonyi} in the original SU(N) gauge theory is played 
by the so called local Polyakov loops, which can be interpreted as static quark
sources with respect to the considered gauge theory.

The study of the deconfinement PT as a percolation process of clusters was 
advocated and started in \cite{Satz:I,Satz:Ib}.  
Recently this direction of research was continued \cite{Gatt:I, Gatt:II}.
Despite many interesting findings, the authors of \cite{Gatt:I, Gatt:II} 
mainly concentrated on the properties of the largest and second largest 
3-dimensional  clusters formed by same-sign Polyakov loops and, hence, they did not 
pay  attention to the properties of the multi-cluster structures made by  smaller 
clusters. However, since the Fisher droplet model \cite{Fisher-67,Fisher-69} 
has been formulated, it became clear that there is no principal difference 
between the critical properties of liquid droplets in systems  
experiencing a liquid-gas PT on one hand and the properties of large spin 
domains in symmetric spin systems with a magnetization PT on the other hand. 
Note that exactly this similarity is able to naturally explain the fact 
that the critical exponents of ordinary simple liquids are the same as the 
critical indices of symmetric spin systems \cite{Fisher-67,Stanley:99,RGmethod}.

Further exploration of the properties of physical clusters has led to the 
formulation of several successful cluster models showing a first order 
PT \cite{Bondorf,Cluster:1,Cluster:2,Bugaev_00}. The Statistical 
Multifragmentation Model of atomic nuclei \cite{Bondorf} treats the 
nuclear fragments of all sizes as charged droplets with a temperature 
dependent surface tension. Its exactly solvable 
version \cite{Bugaev_00} differs from the Fisher model \cite{Fisher-67} 
mainly by the  way of accounting  for  a hard-core repulsion  
between the nuclear fragments. It was a great surprise to learn that these 
two apparently similar models belong to  different universality 
classes \cite{Reuter_01}. Also the Fisher droplet model \cite{Fisher-67} 
has been extended further in order to better reproduce the properties of real 
liquids \cite{Cluster:1,Cluster:2}.

Another interesting development was made in  \cite{GasOfBags:81}, where 
the ``gas of bag model''  was formulated. In contrast to the usual 
cluster models \cite{Fisher-67,Bondorf,Cluster:1,Cluster:2,Bugaev_00},  
it did not take into account the surface tension of quark-gluon bags.  
Two  generalizations  
of the ``gas of bag model''    which incorporate the surface tension for  large 
quark-gluon bags \cite{Bugaev_07,Bugaev_09} were proposed and solved exactly. 
An inclusion of a surface tension
which vanishes at certain values $T_c$ of temperature and $\mu^B_c$ of
baryonic chemical potential has allowed one to formulate statistical 
models with a first order deconfinement PT and/or a cross-over, between 
which there exists either a tricritical \cite{Bugaev_07} or a critical 
endpoint \cite{Bugaev_09} located at  $(T_c,\mu^B_c)$. 
Actually, these models have naturally suggested a physical reason
responsible for the degeneration of a first or second order PT into a 
cross-over. In \cite{Bugaev_07} it was 
demonstrated that in the cross-over region the surface tension becomes 
negative and this very fact prevents the 
formation of the infinite cluster which otherwise would represent the liquid 
phase in all cluster models with first order PT. Because of the important 
role played by the surface tension, these models \cite{Bugaev_07,Bugaev_09}  
were named as ``quark-gluon bags with surface tension'' (QGBST).

Despite the success of the QGBST models in describing the critical 
exponents of some universality classes of spin systems such as the O(2)-O(4) 
model \cite{Ivanytskyi,Ivanytskyi2} and  a novel universality class
called ``the non-Fisher universality class'' \cite{Ivanytskyi3}, 
it is clear that these phenomenological models are dealing with an 
oversimplified picture of quark-gluon bags.

{
Traditionally, in lattice QCD the deconfining PT  is considered as a break down of discrete symmetry of a certain color group.  In our opinion, however,  such a language is  more suited to study the liquid-solid or solid-solid PTs.  From the heavy ion collision experiments we know that   at low energy density the hadron phase is a mixture of gases \cite{Stock12}, while  at  high energy densities achieved 
at modern colliders the QGP is the most perfect liquid  (see e.g.  \cite{Shuryak:08}).
Therefore,    it would be more appropriate to consider the deconfinement PT  as the liquid-gas one. 
 On the other hand, from the discussion above it  is also clear that without the essential  input from lattice QCD a  further progress 
 in developing 
 cluster models for  the deconfinement PT    is simply impossible. }

Therefore, in this work we would like to study the deconfinement PT in 
SU(2) gluodynamics 
on a finite lattice  on the basis of 
 cluster models \cite{Fisher-67, Bugaev_00,Bugaev_07,Bugaev_09,Ivanytskyi,Ivanytskyi2,Ivanytskyi3,Bugaev:05,Bugaev:07} which describe  the liquid-gas PT. 
For this purpose we will identify the geometrical clusters (a \' la Gattringer)
formed from the local Polyakov loops (continuously valued spins)  
and study such important 
properties of the system as the value of the Fisher topological exponent $\tau$ and the 
temperature dependences of  the reduced surface tension 
$\sigma_A$  and of
the reduced chemical potential $\mu_A$
which  is usually 
generated in finite systems  by the  interaction between clusters \cite{Bugaev:05,Bugaev:07}.
{Also we would like to clarify the question what  is  the internal structure of the largest
cluster which  we consider as the gluon bag.}

We consider the SU(2) gluodynamics on a finite lattice as a 
theoretical laboratory to study a well-known second order PT in a finite 
system with strong interaction and  as a testing ground 
to verify  the validity of the  existing  statistical models of cluster type to describe a non-Abelian gauge 
model.  
We restrict ourselves 
for this explorative study
to  SU(2) gluodynamics because this is the simplest and very well known non-Abelian gauge 
model and, hence, we hope to study in details the question on how the Z(2) symmetry breaking is reflected in changes  of  the collective properties  of  
 geometrical clusters formed by Polyakov loops. As it is shown below  such an approach allowed us to introduce new order parameters 
and relate them to the mean value of Polyakov loop.  
Hence, we consider this  work as the first  attempt to  reformulate  lattice QCD 
thermodynamics into the language of statistical cluster models. 

The work is organized as follows. In Section II we give the necessary 
definitions and recall the basic assumptions of the liquid droplet model. 
Section III is devoted to the discussion of the 4- and 3-parametric fits 
of the  size distributions in  the two types of cluster gases formed by 
the Polyakov loops. Such a fit allows us to determine the reduced surface tension 
of the clusters. The properties of the physical surface tension of the two 
types of clusters are discussed in Section IV. In Section V  we calculate 
several critical exponents and explicitly demonstrate how the reduced surface 
tension coefficient can be used as an order parameter. 
Our conclusions are collected in Section VI. 

\section{A liquid droplet model parameterization for SU(2) gluodynamics}
We have performed simulations of SU(2)  on the fixed lattice with 4-dimensional volume 
$N_\sigma^3 \times N_\tau$, where $N_\sigma=24$ and $N_\tau=8$. 
Similarly to \cite{Gatt:I,Gatt:II} we introduce a cut-off value for the 
Polyakov loops $L_{cut} > 0$, whose convenience will be discussed later.  
The Polyakov loops were evaluated  at each spatial base point of the 
4-dimensional lattice. 
The lattice data were generated for 13  values of  inverse coupling
constant $\beta = \frac{4}{g^2}$ inside the interval $\beta \in [2.3115; 3]$. 
Here  $g^2$ is the lattice QCD coupling constant. Since
the lattice spacing $a$ changes with $\beta$, i.e. $a(\beta)$, all the physical 
temperatures  defined by 
$T=\frac{1}{N_\tau a(\beta)}$,  are related to the 13 values of  $\beta$ in the investigated  interval.
The density of $\beta$ points was chosen differently in different regions. 
Basically, if there are small changes in the observed distributions,
we used a low density of $\beta$ points, whereas on the 
high-temperature side of the phase transformation we used a higher density 
of $\beta$ points. We emphasize that in contrast to the fixed 
scale approach, where one or few $\beta$ values are chosen and the temperatures 
are changed by changing $N_\tau$, the physical 3-dimensional volume of considered lattice changes 
with the temperature like $V \propto (1/T)^3$.

The main objects of our analysis are the size distributions of  
clusters formed by the Polyakov loops. The local Polyakov loop is real-valued 
for SU(2) gauge theory ranging from $+1$ to $-1$.
For the cluster recognition we identify monomers, dimers, trimers 
and so on, which are built from the neighboring Polyakov loops of the same sign 
(positive or negative, above some threshold).
In contrast to the definitions used in works \cite{Gatt:I,Gatt:II}, 
we prefer to use a terminology which is closer to the one 
used in studies of  spin clusters of various spin 
models \cite{Borg,moretto-03,moretto-05}.
Instead of concentrating on   the cluster of largest size, as in Refs. 
\cite{Gatt:I,Gatt:II}, we analyze the 
clusters of all sizes and signs and treat them as finite liquid droplets.
Since in the case of SU(2) gluodynamics the  Polyakov loops of opposite  signs are
potentially different in their properties,
here we  investigate the size distributions of the clusters of both 
signs in order to elucidate the phenomenon of spontaneous Z(2) symmetry 
breaking. To distinguish them from each other we denote them as anticlusters 
and clusters, as defined later. 

%
\begin{figure}[h]
\centerline{\includegraphics[width=73mm]{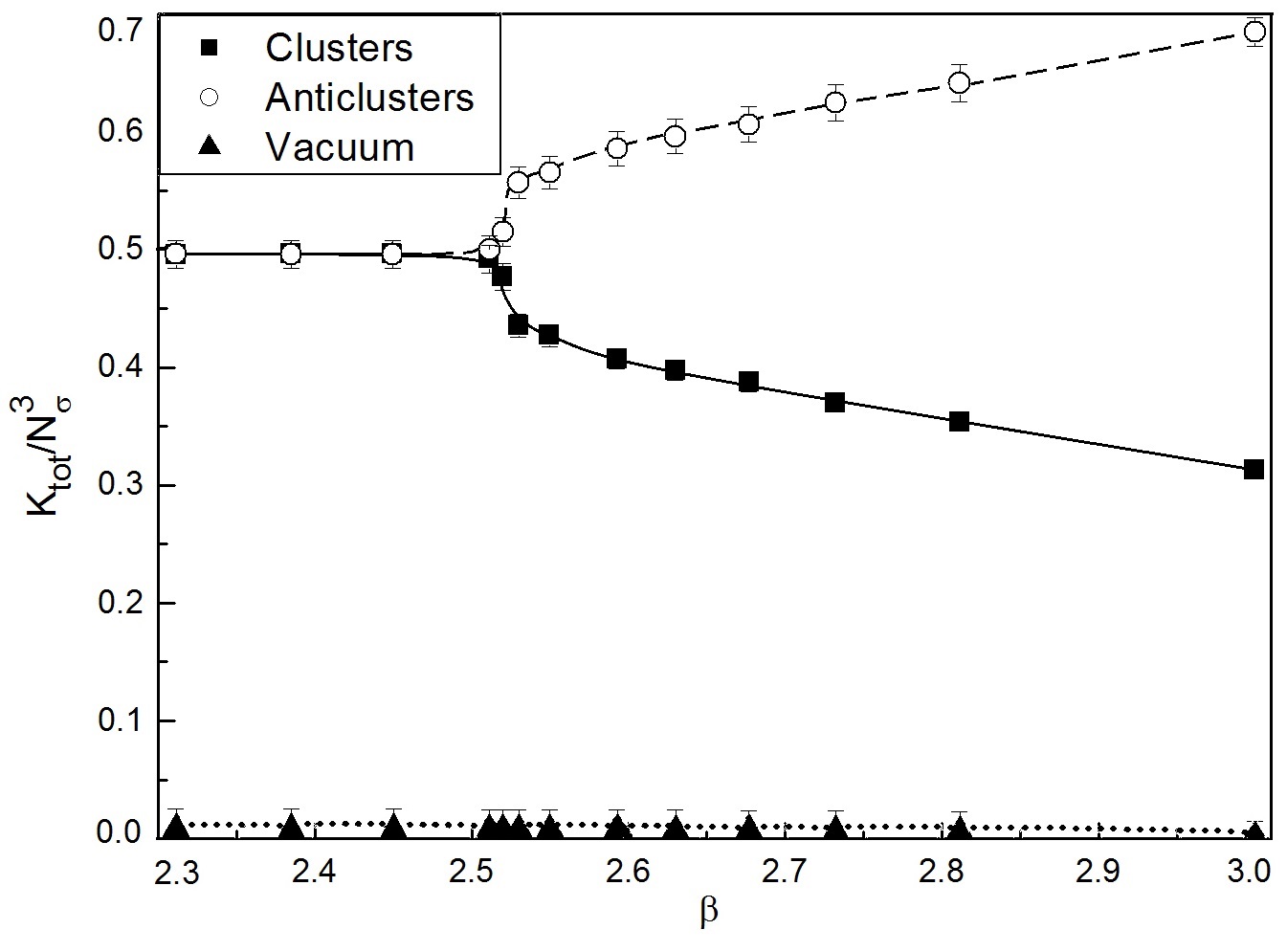} \hspace*{-1.1mm}\includegraphics[width=74mm]{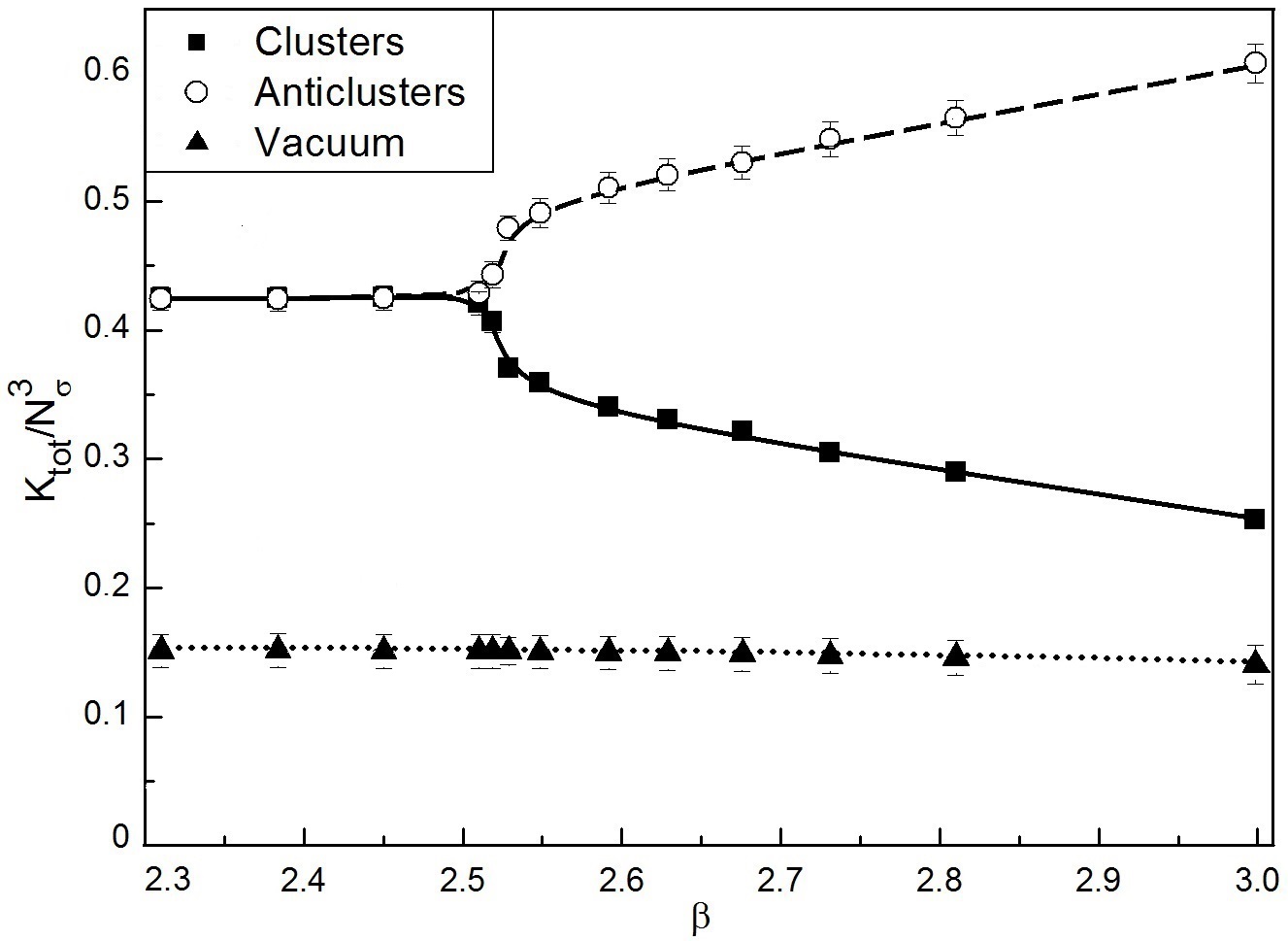}}
\caption{Total volume of clusters, anticlusters and auxiliary vacuum measured 
in units of 3-dimensional  lattice cells $a^3$, relative to the full 3-dimensional  volume, as functions 
of $\beta$. The auxiliary vacuum is independent of $\beta$. Results are shown 
for $L_{cut} =0.1$ (left panel) and for $L_{cut} =0.2$ (right panel). 
The curves are shown to guide the eye. 
}
\label{fig1}
\end{figure}

Let's first recall the formal definition of the local Polyakov loop. 
Traditionally, the volume average of the Polyakov loop is used as an order 
parameter in pure gauge theories. The local Polyakov loop $L(\vec x)$, located 
at a spatial point with the coordinates $\vec x$, is defined by the trace of 
the product of temporal gauge links $U_4 (\vec x, t)$
\begin{eqnarray}\label{EqI}
L(\vec x) = Tr  \prod\limits_{t=0}^{N_\tau-1} U_4 (\vec x, t) \,.
\end{eqnarray}
This expression shows that a local Polyakov loop is a gauge transporter which 
can be considered as the propagator of an infinitely heavy static quark at 
position $\vec x$ forward in time $t$. An important aspect is that the 
ensemble average of the spatially averaged Polyakov loop values is related
to the free energy $F_q$ of a single heavy quark as 
\begin{equation}\label{F/T}
\Big< \frac{1}{V} \int d^3{\vec x} L(\vec x) \Big> \simeq \exp(-F_q/T) \, ,
\end{equation}
where $T$ is the physical temperature of the system. 

Similarly to Refs. \cite{Gatt:I,Gatt:II}, the local Polyakov loop $L(\vec x)$ 
located in a 3-dimensional  point with coordinates $\vec x$ is counted as ``spin up'', 
if $L(\vec x) > L_{cut}$. Analogously, a Polyakov loop $L(\vec x)$ is counted 
as ``spin down'', if $L(\vec x) < - L_{cut}$, whereas the Polyakov loops with 
intermediate values $- L_{cut} \le L(\vec x) \le  L_{cut} $ are not accounted 
in the analysis and are regarded as an ``auxiliary'' (``confining'') vacuum. 
Such a definition is convenient, since for a given cut-off the volume fraction 
of the auxiliary vacuum is independent of $\beta$ value,  as one can see from 
Fig. \ref{fig1}.
A given ``spin up'' is considered as a monomer, if all its nearest neighbors are 
either ``spin down'' or belong to the auxiliary vacuum. Similarly, two neighboring 
``spins up'' form a dimer, if all their nearest neighbors are either ``spin down'' 
or belong to the auxiliary vacuum. 
In the same way one can define arbitrarily large $n$-mer for spins up and down. 
Besides a formal convenience, the cut-off $L_{cut}$ plays an important role 
in studies of the thermodynamic limit on the lattice \cite{Gatt:II}.

\begin{figure}[b]
\centerline{\includegraphics[width=72mm]{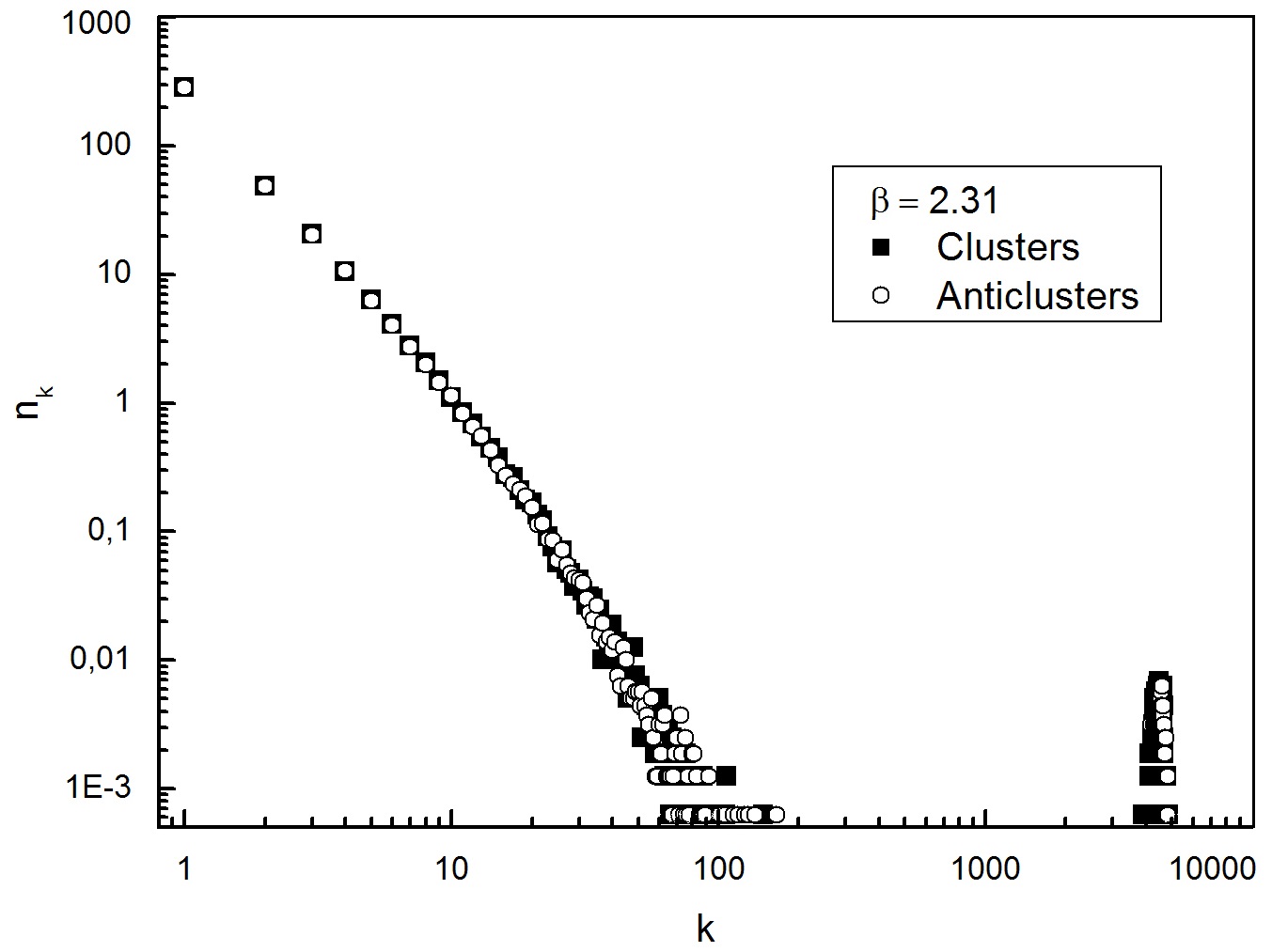} \hspace*{1.1mm}\includegraphics[width=72mm]{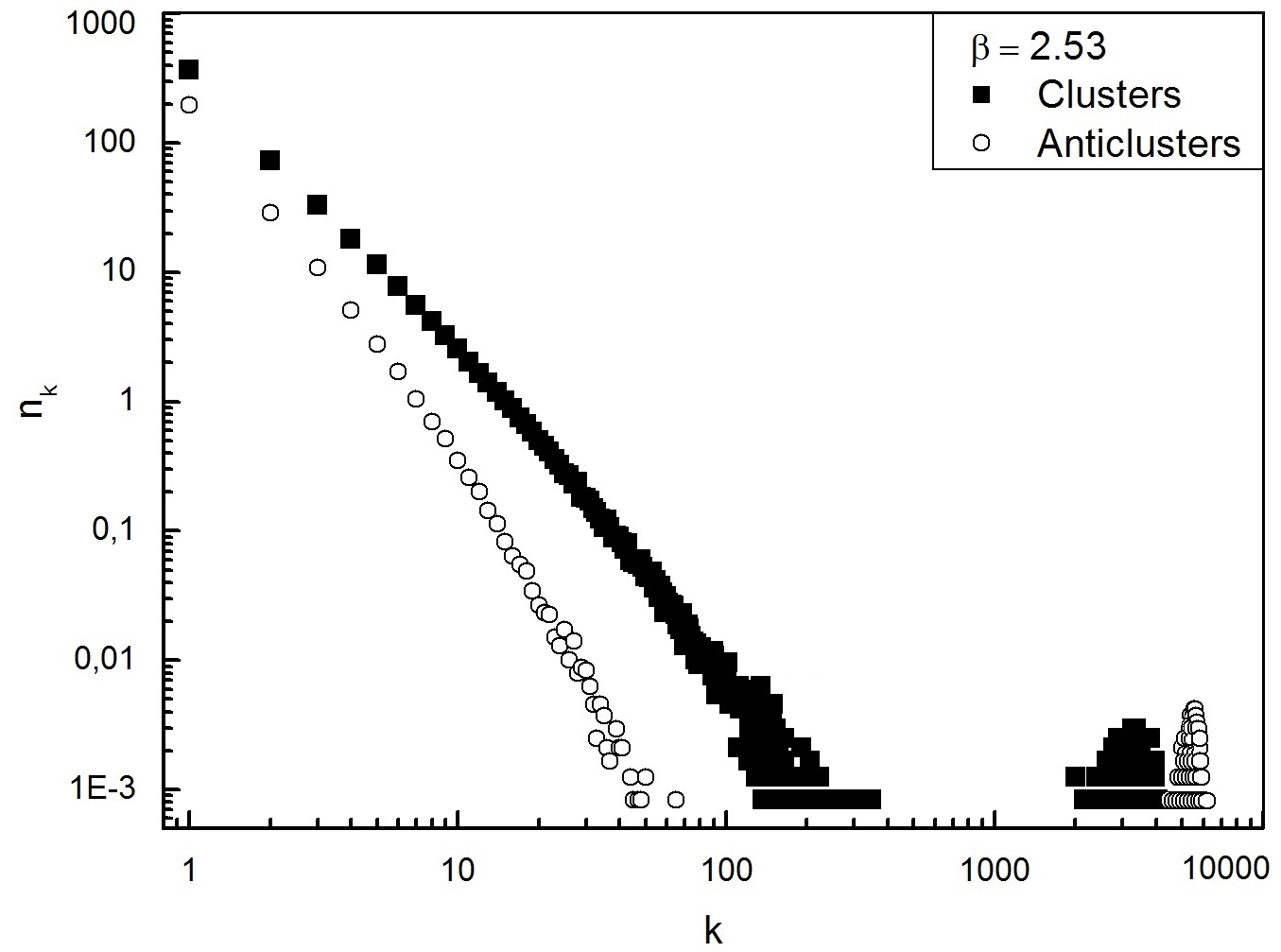}}
\vspace*{1.1mm}
\centerline{\includegraphics[width=73mm]{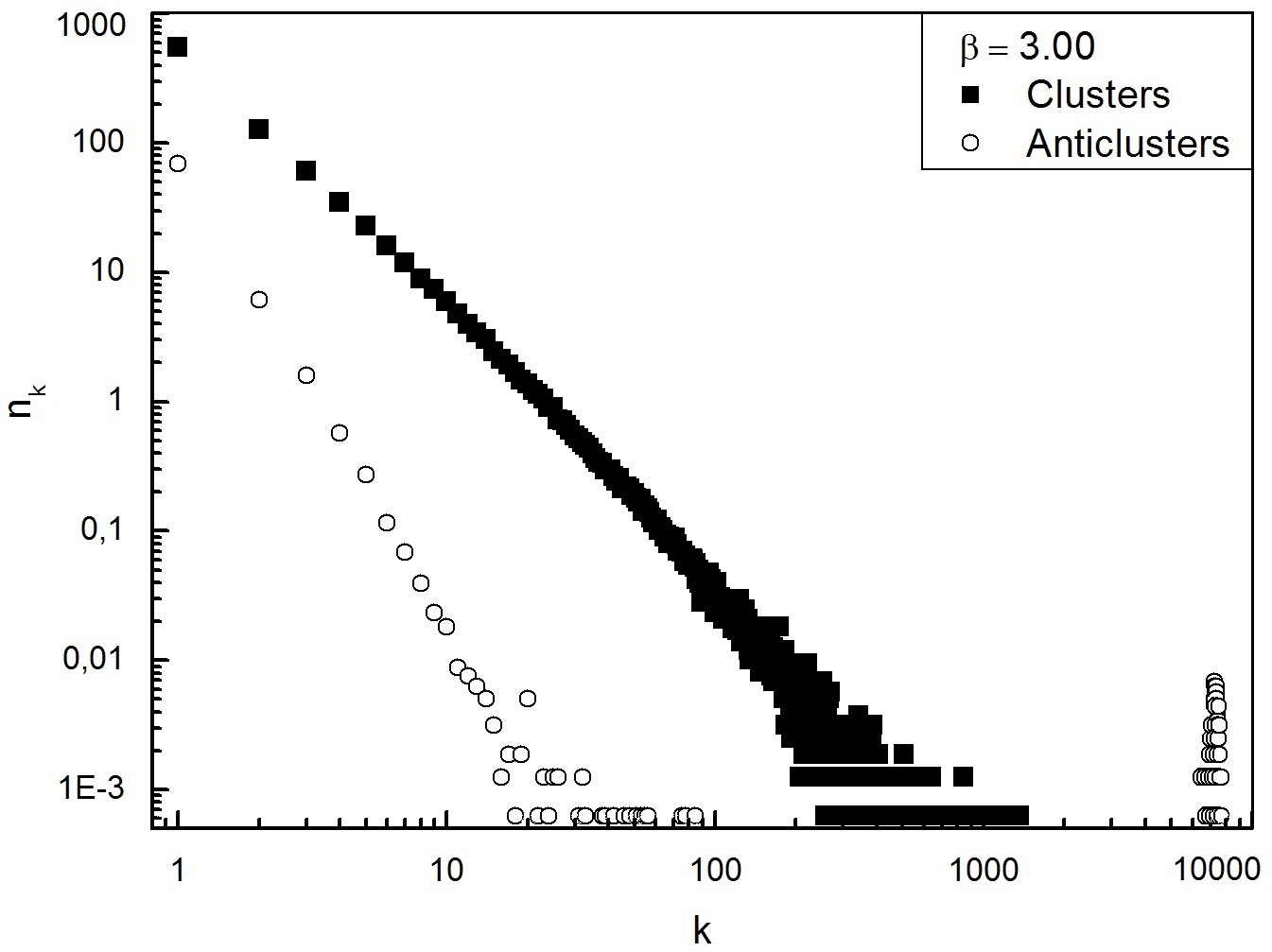}}
\vspace*{-4.2mm}
\caption{Ensemble average  size distributions of clusters (squares) and anticlusters 
(circles) for different $\beta$ values. $k$ is the size of an (anti)cluster,
$n_k$ is its multiplicity. 
{\bf Upper left panel:} For $\beta < \beta_c^\infty $ 
($\beta_c^\infty = 2.5115$ is the deconfinement point for $N_\tau=8$ in the gauge system) 
one can see a complete symmetry between the two distributions. 
The largest (anti)cluster is of the mean size $k \simeq 4500$, which  is well separated from the distribution 
of  the corresponding gas of smaller (anti)clusters.
{\bf Upper right panel:} Same as in the upper left panel, but for 
$\beta = 2.53$ which is slightly above $\beta_c^\infty$. One can 
see now an emerging difference between the two distributions:
the multiplicities in the gas of clusters are higher than the multiplicities
in the gas of anticlusters, whereas the largest anticluster is noticeably 
larger than the largest cluster. 
{\bf Lower panel:}  Same as in upper panels, but for $\beta=3.0$, which is well 
above $\beta_c^\infty$. The largest cluster is now inseparable from the 
corresponding cluster gas. 
The statistical errors are of the order of the size of symbols or smaller. 
All these results refer to a cut-off $L_{cut}=0.2$.
}
\label{fig2}
\end{figure}

Let us call the largest $n$-mer on the lattice as the ``anticluster droplet'', 
whereas all other $n$-mers of the same spin sign (being smaller in size) are 
considered as the ``gas of anticlusters''. Similarly, let's call the largest 
$n$-mer of the opposite spin as a ``cluster droplet'', whereas other clusters 
of the same sign (which are smaller in size) are considered as a ``gas of 
clusters''. As one can see from Fig. \ref{fig2}, these definitions are 
sensible not only for each generated configuration of gauge fields, but 
also for the ensemble average over many configurations, because in most 
cases (except for highest values of $\beta$) the largest (anti)clusters 
are well separated from the corresponding gases. 
From Fig. \ref{fig1} one can conclude that for a given cut-off the sum of  
cluster and anticluster volume fractions is $\beta$-independent and 
consequently the volume fraction of auxiliary vacuum is $\beta$-independent,
too.

The average distributions $n_k$ shown in Fig. \ref{fig2} for the set of 
$\beta$-values have been calculated for ensembles of 800 and 1600 or 2400 
independent configurations of gauge fields on the lattice. If the individual 
distributions found for 800 and 1600 independent configurations did not differ 
from each other within the statistical error bars, then we used the distribution 
averaged over 1600 configurations as high-statistics limit. 
In most cases such distributions have  converged in the described sense, whereas
for $\beta= 2.52$, $\beta= 2.53$ and $\beta= 2.677$ we were forced to enlarge the 
ensemble to 2400 independent configurations to calculate the averaged distributions. 
In fact, the configurations under analysis are separated by 10 Monte Carlo sweeps.

The size distributions of Fig. \ref{fig2} are very similar to the ones found 
for 2- and 3-dimensional Ising systems  \cite{Borg,moretto-03,moretto-05} and to the nuclear 
fragment size distributions inside the mixed phase of a first order PT obtained 
within  the statistical multifragmentation 
model \cite{sagun-14} (see Figs. 5 and 6 therein).  
As one can see from Fig. \ref{fig2}, the typical structure of the size 
distributions of Polyakov loop clusters contains the following features:
\begin{enumerate}
\item A branch with average multiplicities monotonously decreasing with the cluster size  which can be associated with the gas of clusters (and similarly for 
anticlusters) 
\cite{Bondorf,THill:1,Moretto:97,DGross:1,Bmodal:Chomaz01,Bmodal:Lopez06,moretto-05}.
\item The largest cluster (anticluster) which is separated from the corresponding gas 
distribution by a gap and which, according to the liquid-gas PT phenomenology, 
can be associated with the corresponding liquid
\cite{Bondorf,THill:1,Moretto:97,DGross:1,Bmodal:Chomaz01,Bmodal:Lopez06,moretto-05}.  
The distribution represents the fluctuations in size of largest (anti)cluster in different events. 
\item The gap always exists for anticlusters, whereas for clusters 
it gradually becomes narrower for $\beta > 2.51$ while simultaneously 
the size distribution of the largest cluster evolves towards smaller sizes 
in such a way that at $\beta \simeq 3.0$ it gets inseparable from the 
size distribution characterizing the gas of clusters.
\end{enumerate}

Based on the similarity of size distributions of Polyakov loop (anti)clusters 
and the ones of the Ising spin clusters and nuclear fragments, we decided to 
inquire whether the liquid droplet model (LDM) formula \cite{Fisher-67,Bugaev_00}
        \begin{equation}\label{EqIb}
                 n^{th}_A (k) = C_A\, \exp\left(\mu_A k-\sigma_A k^\varkappa-\tau_A \ln k\right) \,,
        \end{equation}
is able to reproduce the distributions of clusters (A=cl) and anticlusters (A=acl) 
found in the course of the Polyakov loop cluster analysis.  
As usual in cluster 
models \cite{Fisher-67,Cluster:1,Cluster:2,Bugaev_00,Reuter_01,Ivanytskyi,Ivanytskyi2,Ivanytskyi3}  
the first term in the exponential on the right hand side of (\ref{EqIb}) 
defines the bulk free energy of $k$-volume (anti)cluster, the second term 
corresponds to a surface free energy of such (anti)cluster, while the last 
term represents  the Fisher term with critical exponent $\tau_A$. 
Fitting the (anti)cluster distributions, we can determine their reduced 
chemical potential  $\mu_A$ (in units of temperature), the reduced surface tension 
coefficient $\sigma_A$ (in units of temperature), the Fisher exponent 
$\tau_A$ and the normalization factor $C_A$. 
The true chemical potential and surface tensions are, respectively, defined as 
$T \mu_A$ and  $T \sigma_A$.

Similarly to the 3-dimensional cluster 
models \cite{Fisher-67,Bondorf,Cluster:1,Cluster:2,Bugaev_00,Reuter_01,Ivanytskyi,Ivanytskyi2,Ivanytskyi3}, 
the power $\varkappa$ relating the mean surface and mean volume of (anti)clusters 
was fixed by dimensionality to $\varkappa = \frac{2}{3}$. 
Fixing the value of the power $\varkappa$ in this way, we reduce the number 
of fitting parameters to 4 for each type of clusters. These parameters are 
$C_A$, $\mu_A$, $\sigma_A$ and $\tau_A$  with A=\{cl, acl\}.
We are perfectly aware about the fractal dimension of large 
(anti)clusters \cite{Gatt:I, Fractals} and this is the reason why we do not 
fit the largest (anti)clusters in the system. 
On the other hand, the  $\chi^2/dof$ values obtained by a four parametric fit 
are about 1 and this means that to leading order the effects of fractal 
dimension of the gas of (anti)clusters can be neglected. 

\section{Results of fit and their  interpretation within the LDM-framework}
{\bf Results of  the 4-parametric fit.}
The first question  to be clarified is to determine the minimal 
value of the (anti)cluster size  $k_{min}$, for which 
the LDM parameterization (\ref{EqIb}) is valid. For instance, 
for 2-dimensional Ising clusters the minimal area was found to 
be 10 \cite{moretto-05}, while for 3-dimensional ones it was determined 
to be $k_{min}=3$  \cite{moretto-05}.

The fit procedure corresponds to minimization of $\chi_A^2/dof$ 
        \begin{equation}\label{EqII}
                \frac{ \chi_A^2}{dof}= \frac{1}{k_{max}-k_{min}-4} \sum\limits_{k=k_{min}}^{k_{max}}\frac{\left[ n_A^{th} (k) -n_A(k) \right]^2}{[\delta n_A (k)]^2} \,,
        \end{equation}  
with respect to parameters $C_A$, $\mu_A$, $\sigma_A$ and $\tau_A$,  
independently for clusters (A=cl) and anticlusters (A=acl).  
Here the quantity $\delta n_A (k)$ denotes the statistical error in defining 
the average multiplicity $n_A(k)$ in simulations.

\begin{figure}[ht]
\centerline{\includegraphics[width=73mm]{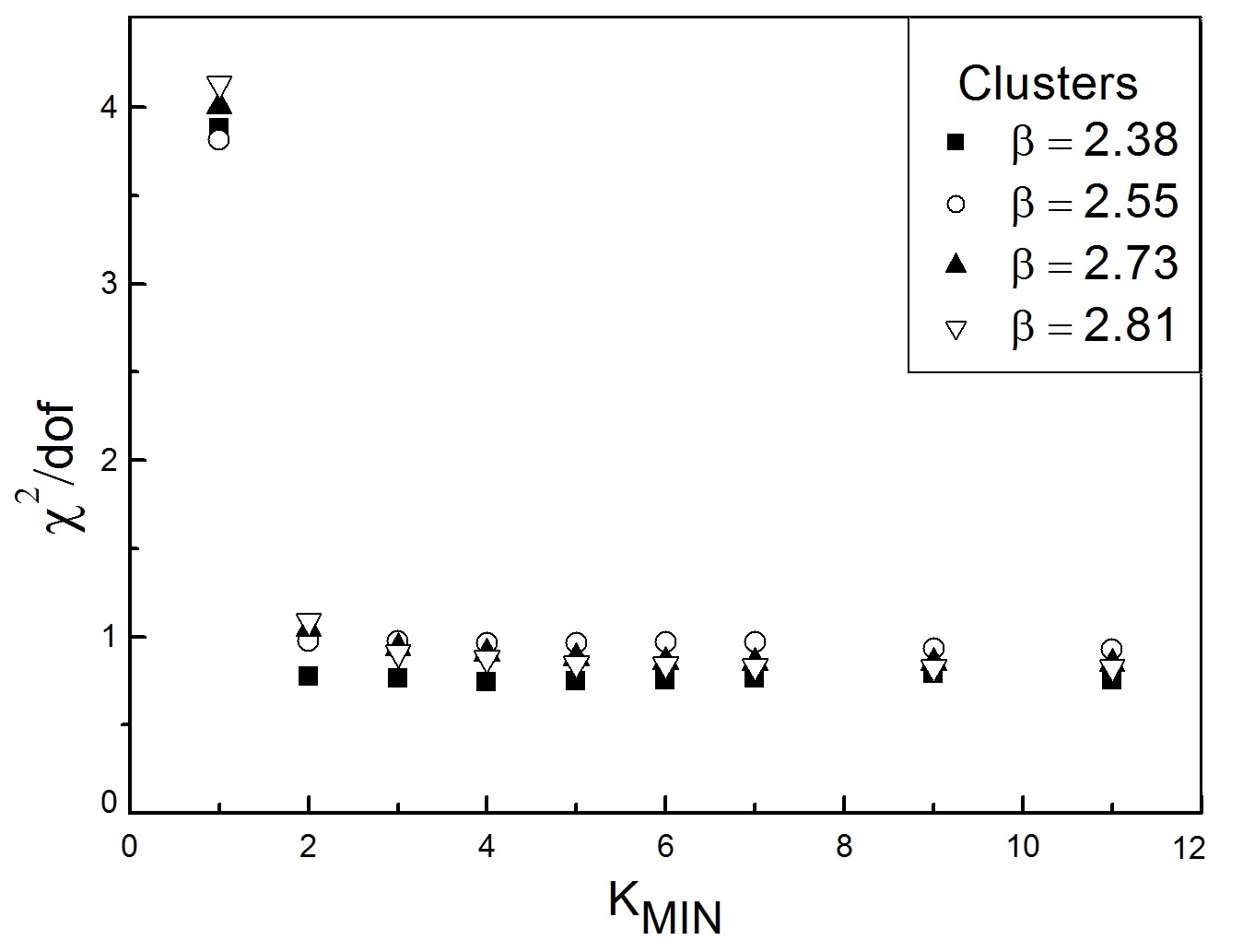} \hspace*{-1.1mm}\includegraphics[width=73mm]{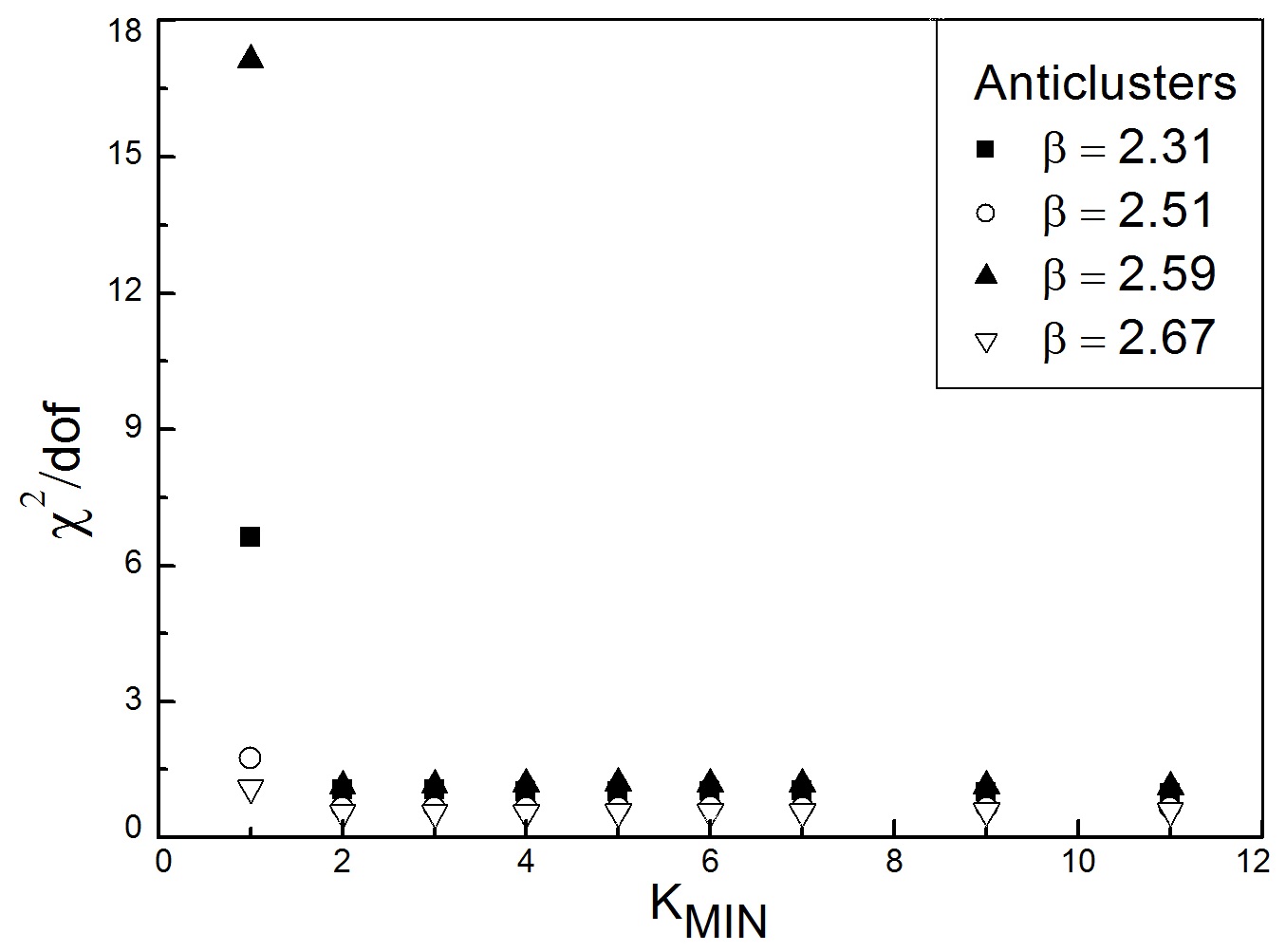}}
\caption{ $\chi^2/dof$ results of the 4-parametric fit by the LDM formula for 
different $k_{min}$ are shown for a few selected values of $\beta$: left for clusters, 
right for anticlusters. 
As one can see from both panels, only the monomers are not described by the 
LDM formula.  Results are shown for $L_{cut} =0.2$.
}
\label{fig3}
\end{figure}

In practice, the minimization of $\chi_A^2/dof$ was done using the gradient 
search method. In other words, if the ``vector of parameters'' 
$\overline{p}_A=(C,\mu,\sigma,\tau)_A$  is known  at some step of iteration 
procedure, then the next approximation is defined as
        \begin{equation}\label{EqIII}
                \overline{p}_A\rightarrow\overline{p}_A-\epsilon \cdot \nabla_{\overline{p}}~\chi_A^2/dof.
        \end{equation} 
Here $\nabla_{\overline{p}}$ is the gradient operator, whereas
$\epsilon=diag(\epsilon_C,\epsilon_\mu,\epsilon_\sigma,\epsilon_\tau)$ 
is a diagonal matrix with positive elements. This simple
scheme corresponds to the steepest descent search for the local extremum 
$\nabla_{\overline{p}}\chi_A^2/dof=0$, where $\chi_A^2/dof$ has a minimum.  
It is necessary to stress that the obtained results demonstrate  a high 
stability with respect to the random variation of the initial values of 
the parameters $\overline{p}_A$. For each minimization search 5 or 6 
different initial values of the parameters $\overline{p}_A$ were taken  
randomly. If all of them led to the same minimum, 
the minimization was stopped. Hence, we are sure that the minima we have 
found are the global ones. 

The upper limit $k_{max}$ of the sum in Eq. (\ref{EqII}) was chosen to account 
for the lattice data with a reliable statistics, i.e. when the statistical 
error of the k-volume (anti)cluster multiplicity $\delta n_A(k)$ is essentially 
smaller than the corresponding multiplicity $n_A(k)$. The maximal volume 
for which such a condition is satisfied 
is denoted as $k_{max}$. The (anti)clusters with the volume larger than $k_{max}$ 
have been excluded from the fit procedure. The value of $k_{max}$ depends on 
$\beta$. For clusters defined with the cutoff $L_{cut}= 0.2$, 
the maximally admitted cluster size increases from $k_{max} \simeq 100$ for 
$\beta= 2.3115$ to  $k_{max} \simeq 300$ for $\beta= 3.0$,
whereas for anticlusters the maximally admitted cluster size gradually
decreases from $k_{max} \simeq 100$ for $\beta= 2.3115$ to $k_{max} \simeq 10-20$ 
for $\beta = 3.0$. 
For clusters defined with the cutoff $L_{cut}= 0.1$ the corresponding range of  
$k_{max}$ for clusters is between $50$ at $\beta= 2.3115$ and $100$ at $\beta= 3.0$, 
while for anticlusters it is between $50$ at $\beta= 2.3115$ and $10-15$ (at most) 
at $\beta= 3.0$ (for more details see below).

\begin{figure}[h]
\centerline{\includegraphics[width=73mm]{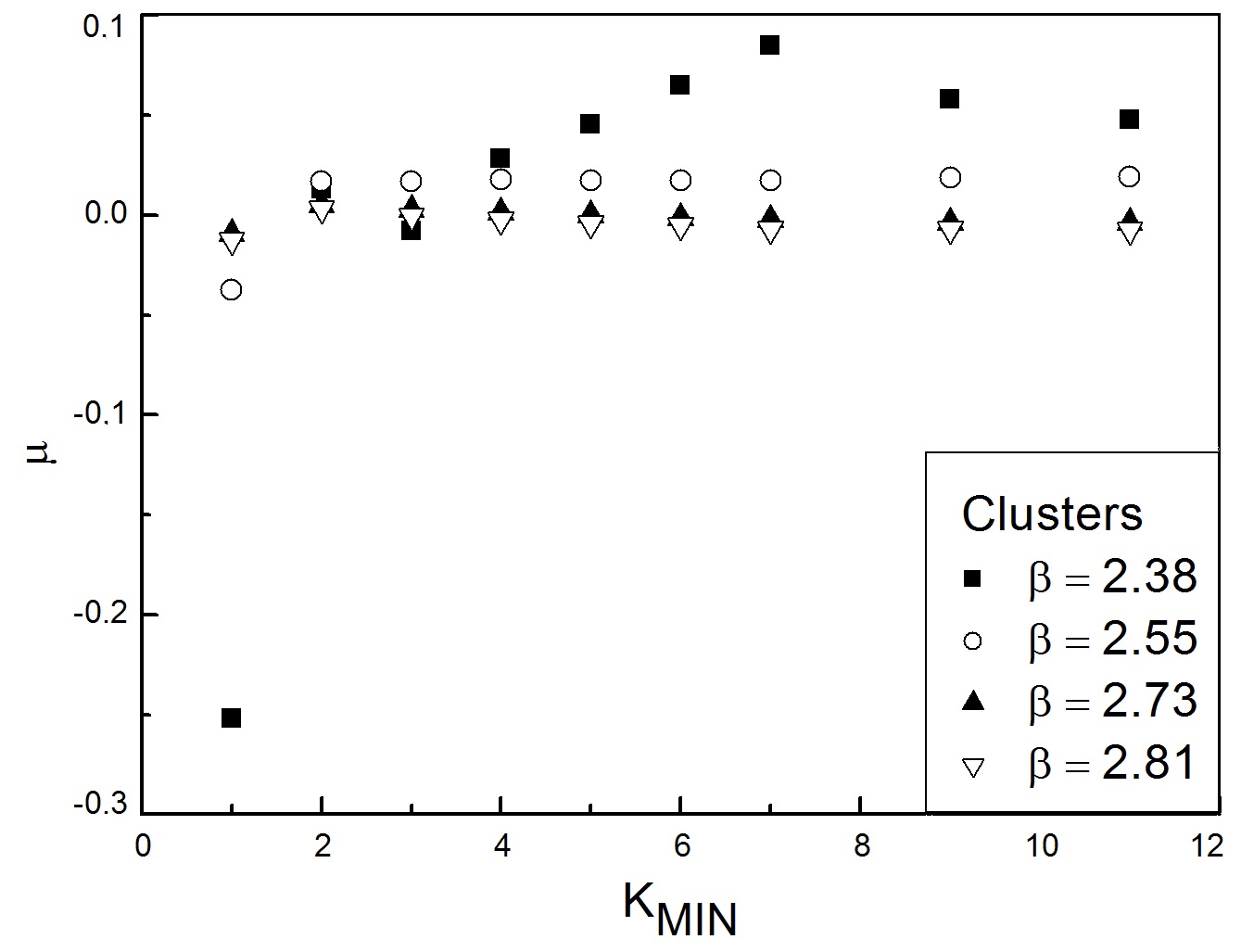} \hspace*{1.1mm}\includegraphics[width=73mm]{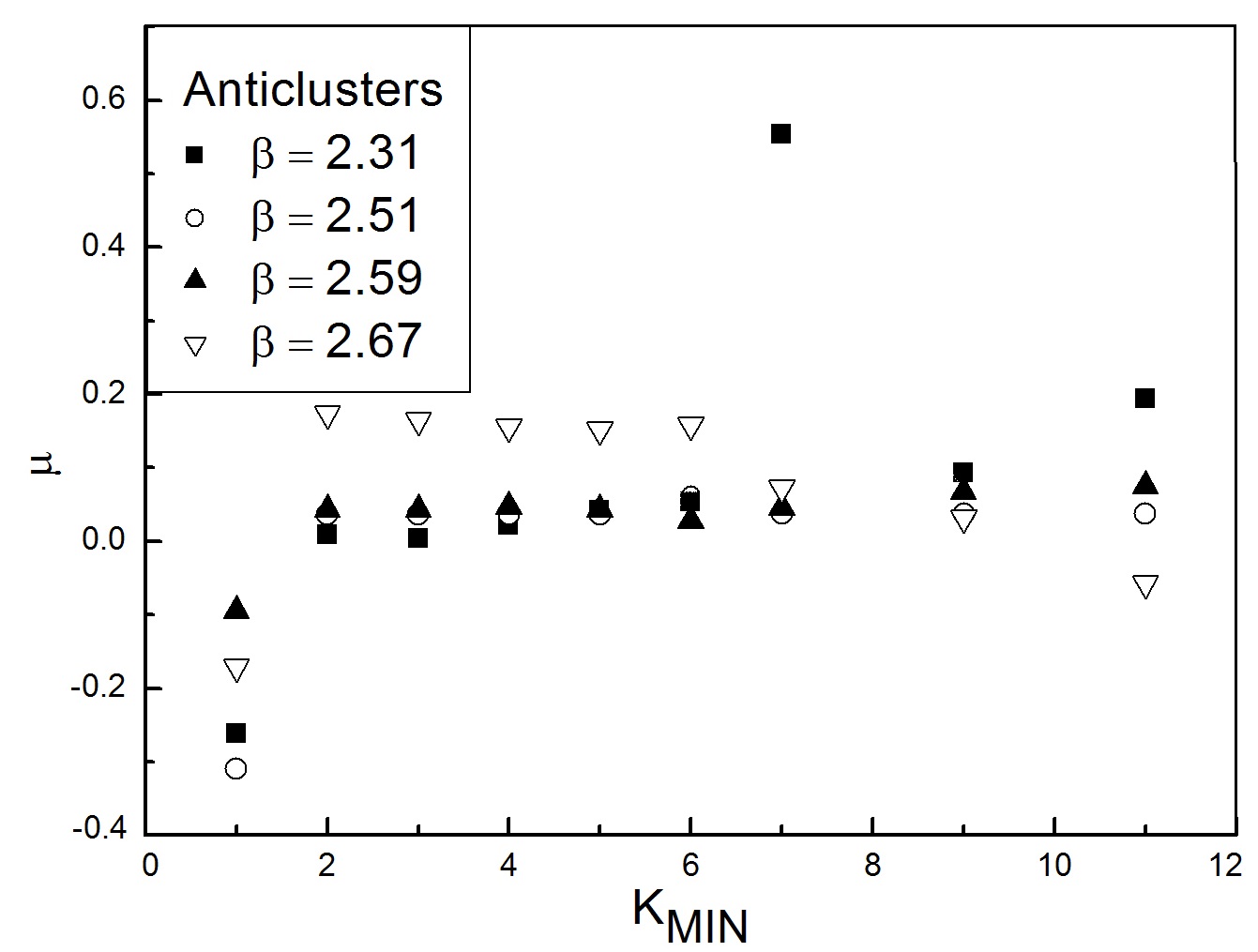}}
\caption{Values of the reduced chemical potentials $\mu_A$, found from 
4-parametric fitting of the LDM formula, are shown for several values 
of $k_{min}$ and for a few values of $\beta$. 
Similar results are found for other values of  $\beta$.
All these results refer to the choice of cut-off $L_{cut}=0.2$.
}
\label{fig4}
\end{figure}

Using the 4-parametric fit procedure for different values assumed for 
the minimal value of (anti)cluster volume $k_{min} \ge 1$, we found
that only monomers ($k_{min} = 1$) are {\it not} described by the LDM formula 
(\ref{EqIb}). The mean deviation squared per number of degrees 
of freedom $\chi^2/dof$  are shown in 
Fig. \ref{fig3} for several values of $\beta$. The results for other values of  
$\beta$ are similar.  
Fig. \ref{fig4} demonstrates the stability of the 4-parametric fit results obtained 
for  $k_{min} = 2-4$ for clusters and 
for  $k_{min} = 2-6$ for anticlusters.

\begin{figure}[h]
\centerline{\includegraphics[width=73mm]{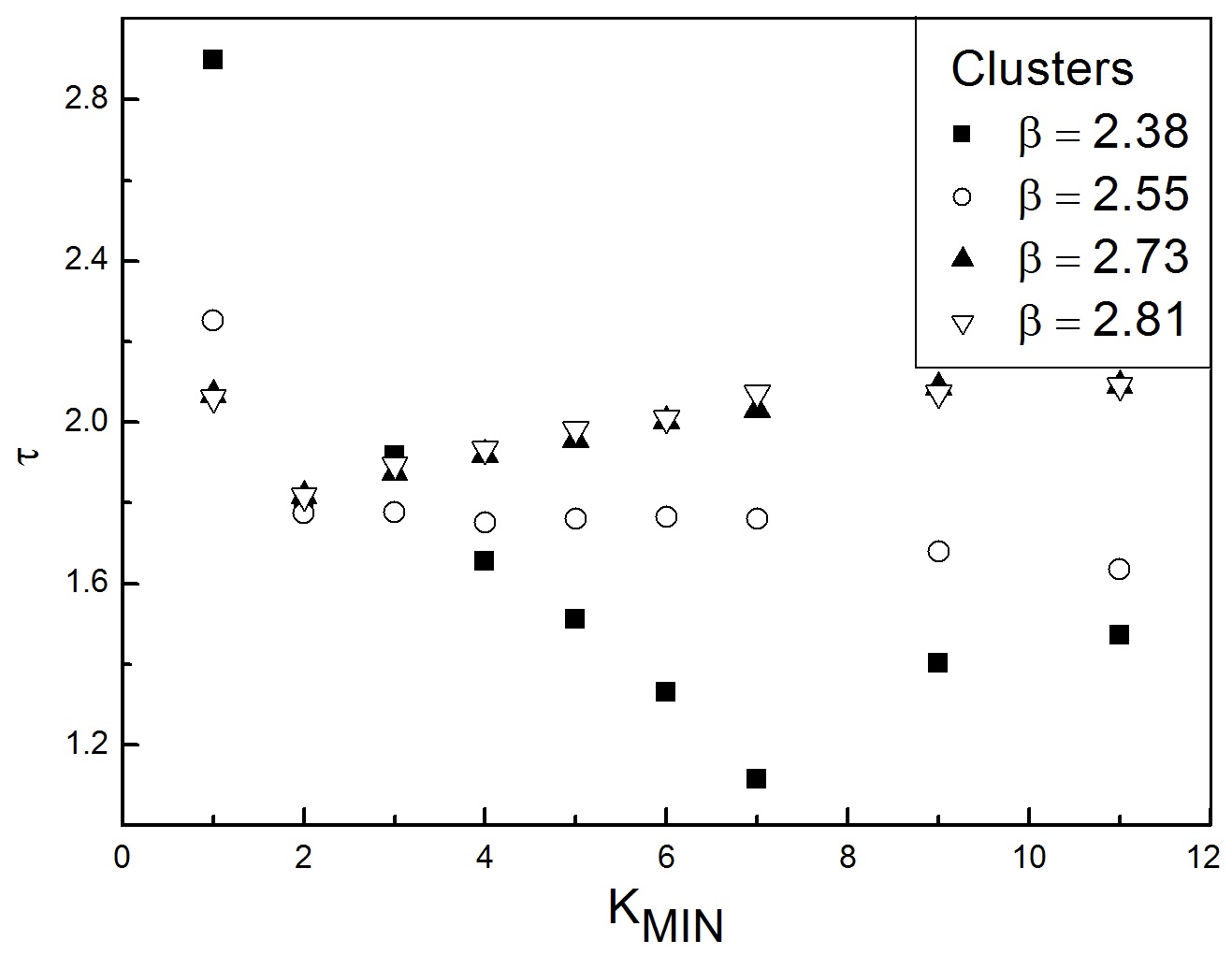} \hspace*{1.1mm}\includegraphics[width=73mm]{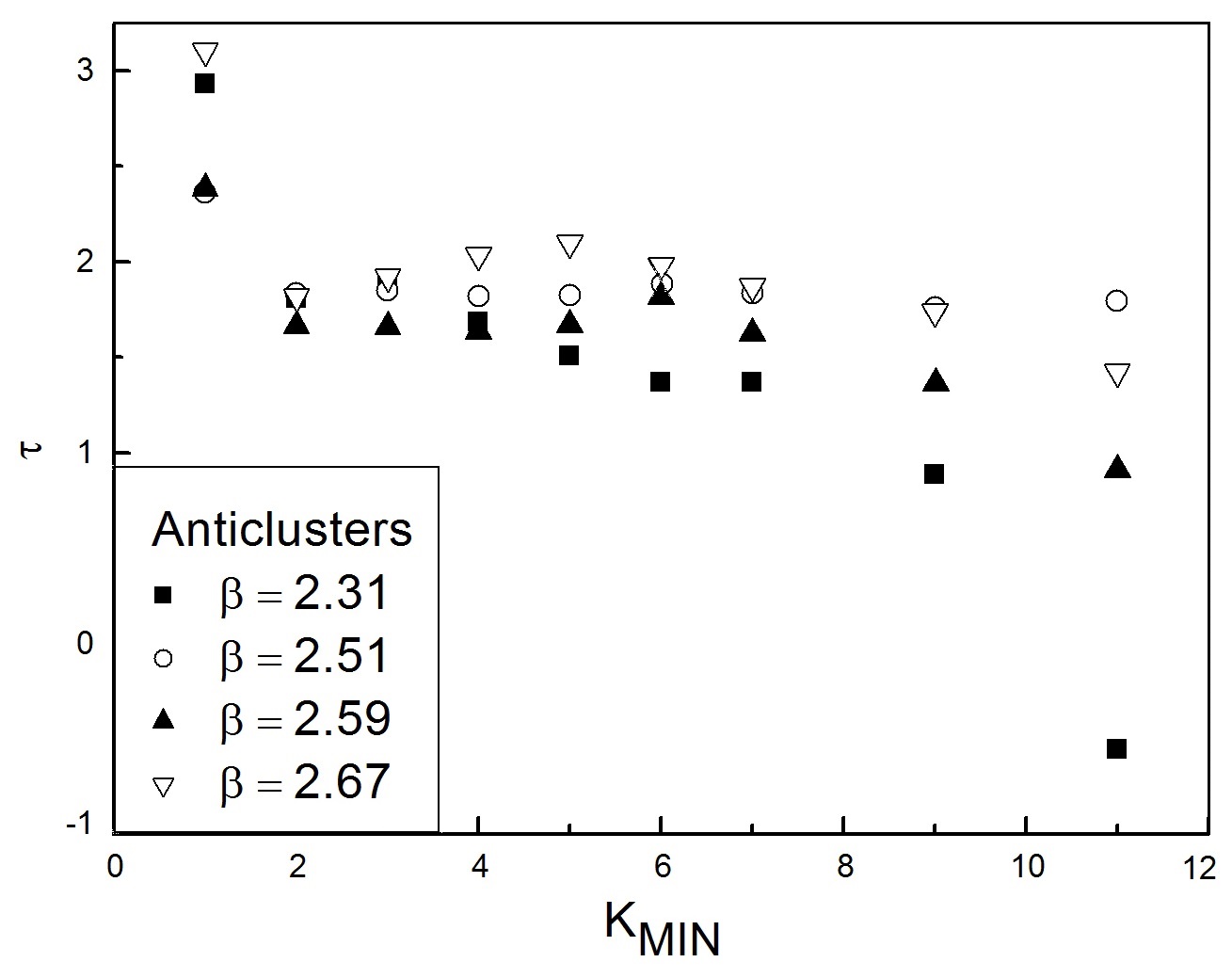}}
\caption{The Fisher exponent $\tau$ for several values of $k_{min}$
and for a few values of $\beta$ found by the 4-parametric fitting 
of the LDM formula. 
Similar results are found for other values of $\beta$.
All these results refer to a cut-off $L_{cut}=0.2$.
}
\label{fig5}
\end{figure}

The 4-parametric fit  allowed us to simultaneously determine 
the $k_{min}$ value  for (anti)clusters and 
the range of values for the Fisher exponent $\tau$. The traditional cluster 
models \cite{Fisher-67,Cluster:1,Cluster:2,Bugaev_00,Reuter_01,Ivanytskyi,Ivanytskyi2,Ivanytskyi3} 
require a temperature independent value for the Fisher parameter $\tau$.  
From Fig. \ref{fig5} one can see that this requirement is well fulfilled both for 
clusters and anticlusters, if $k_{min}=2$ is chosen. 
For $k_{min}=3$ we see that values of  $\tau$  found for different $\beta$ differ 
significantly.
Therefore, we conclude 
that the physically most adequate description of (anti)cluster sets of data 
can be achieved for $k_{min}=2$. This  conclusion is valid for both cut-off 
values studied here.

{
A few details  should be given about  finding the mean value of  the Fisher index $\tau$
and its error. 
For   $k_{min} =2$ we obtain  one  $\beta$ dependent set of the Fisher topological exponents  for clusters and
another set for anticlusters.
Each of these two sets  is characterized by
the average value $\tau_A$ and its dispersion $\delta\tau_A$. It is remarkable that $k_{min}=2$
simultaneously minimizes dispersions $\delta\tau_A$ for  clusters and
anticlusters. 
 Our analysis  shows that 
$\tau_{cl}\simeq1.807$, $\delta\tau_{cl}\simeq0.008$ and $\tau_{acl}\simeq1.756$,
$\delta\tau_{acl}\simeq0.070$.  According to the theory of measurements 
a common value of the Fisher topological exponent  for clusters and anticlusters should be defined as the weighted 
 average with the weights $\omega_A=1/\delta\tau_A^2$, whereas its error  is given by
$\delta\tau=(\omega_{cl}+\omega_{acl})^{-1/2}$ \cite{Taylor82}. Hence, we get
$\tau_A \simeq1.806\pm0.008$. 
}

It is remarkable that in both  these cases, i.e. for clusters 
and anticlusters, we have found $\tau_{cl} <2$ and $\tau_{acl} <2$
as the averaged values for different  $\beta$. 
We believe this is an important finding because in Fisher droplet model 
one 
usually has $\tau > 2$  \cite{Fisher-67,Fisher-69}, namely 
$\tau \simeq 2.07$ for the 2-dimensional case and $\tau \simeq 2.209$ for 
the 3-dimensional model. 
On the other hand, the value $\tau_A \simeq 1.806 \pm0.008$ found here is consistent 
with the results of two exactly solvable cluster models considered in \cite{Reuter_01,Ivanytskyi,Ivanytskyi3}.  
Note, that the question whether $\tau$ is larger or smaller than 2 is of 
principal importance because it determines the universality class of the 
model \cite{Reuter_01,Ivanytskyi,Ivanytskyi2,Ivanytskyi3}.

\begin{figure}[h]
\centerline{\includegraphics[width=73mm]{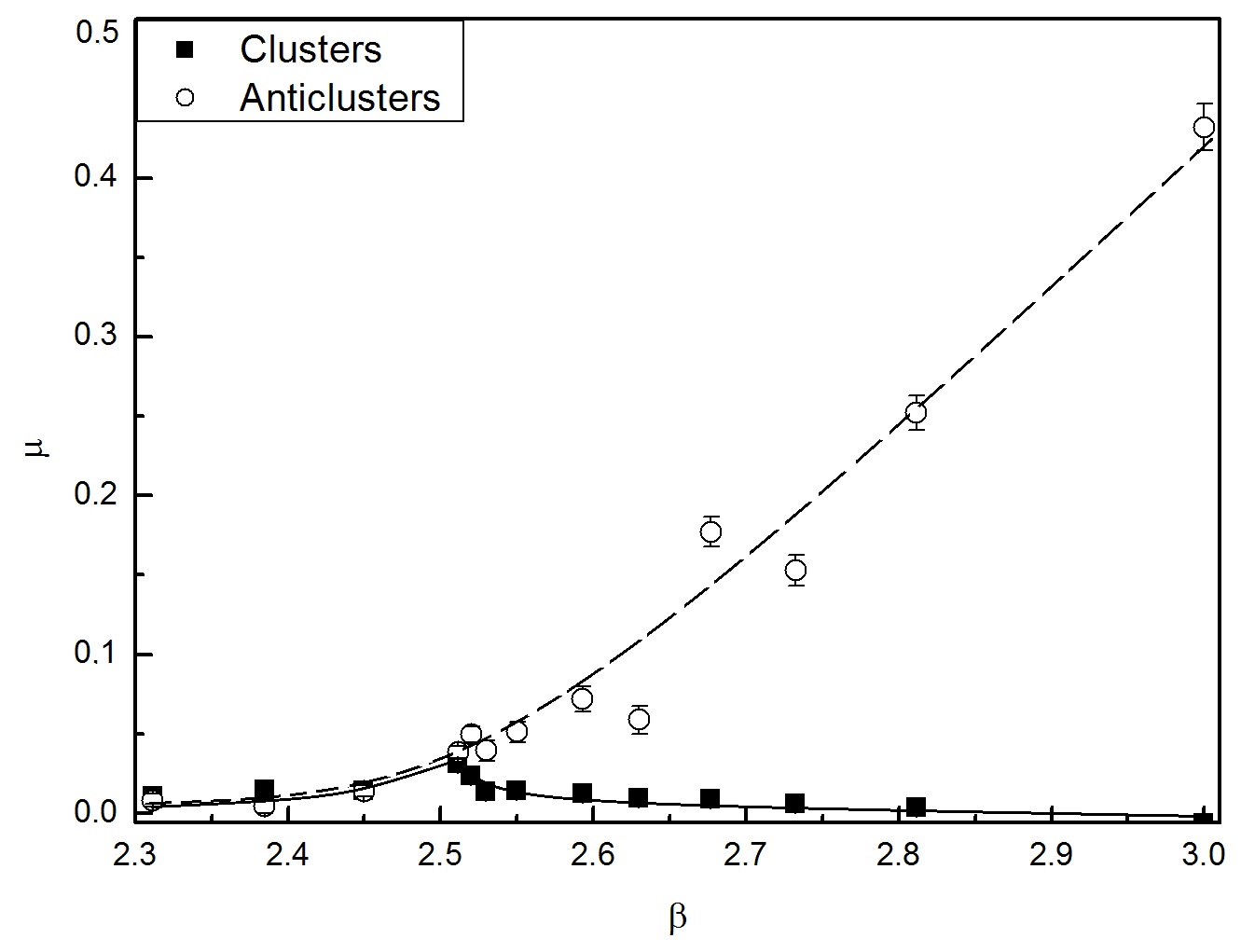} \hspace*{1.1mm}\includegraphics[width=73mm]{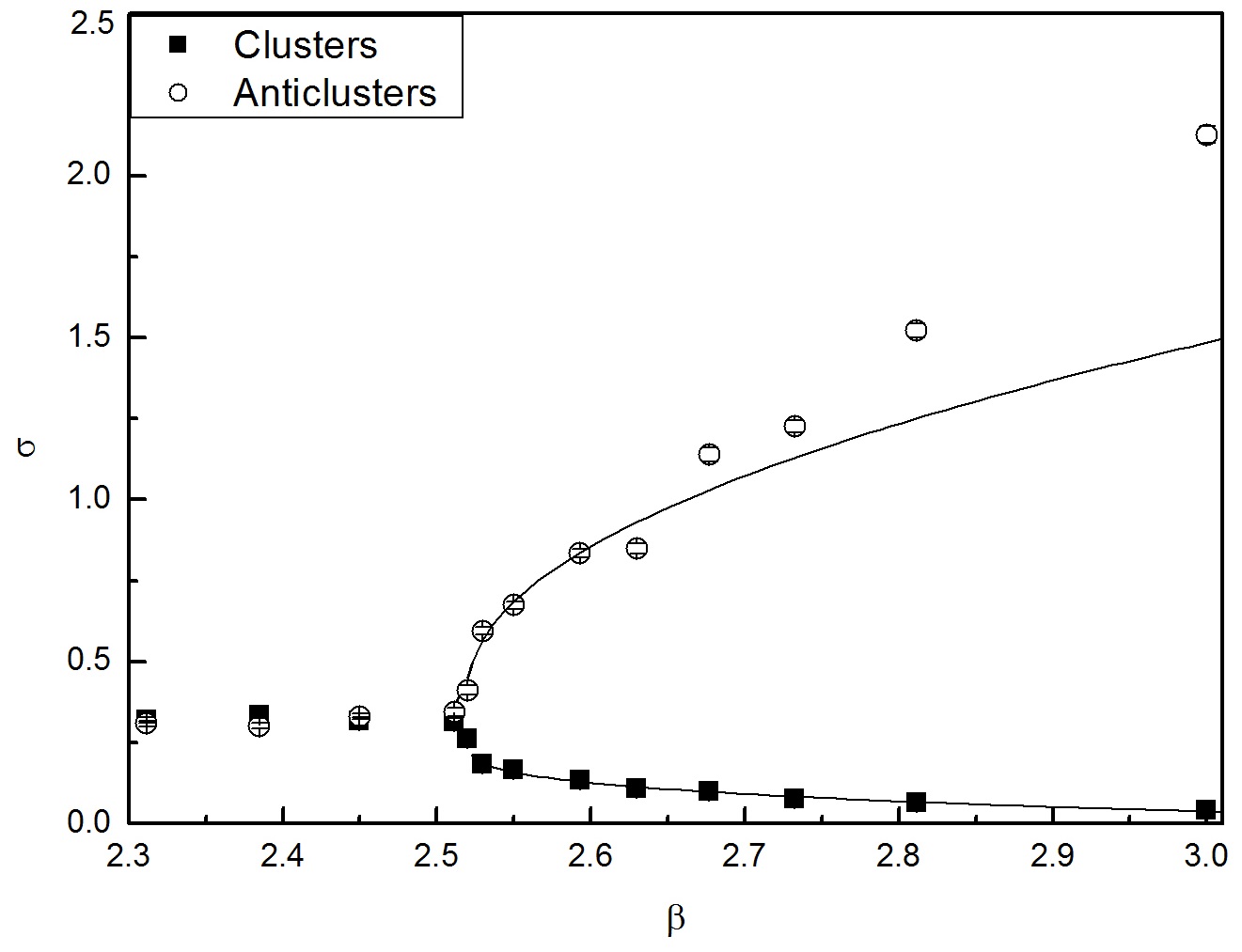}}
\vspace*{1.1mm}
\centerline{\includegraphics[width=73mm]{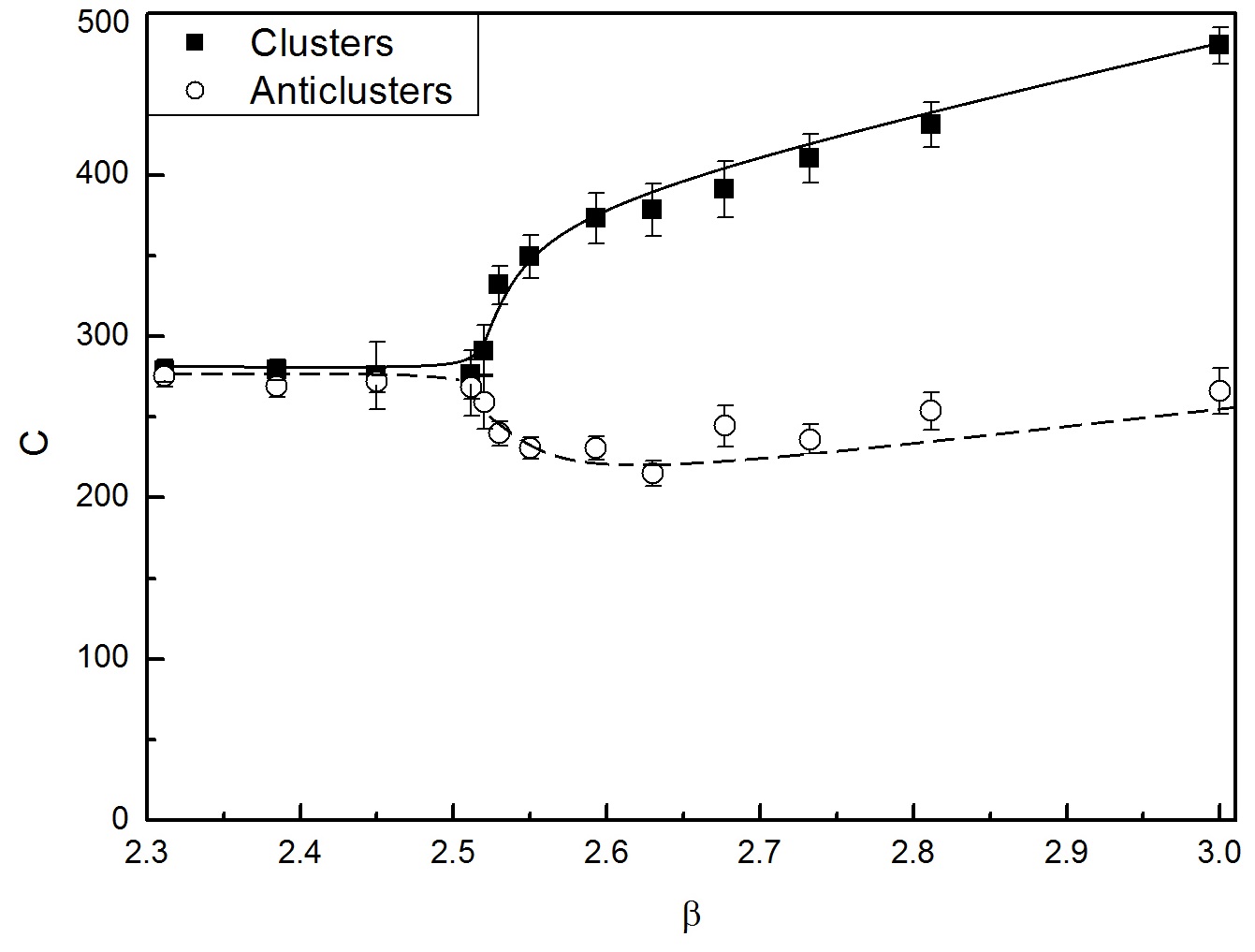} \hspace*{1.1mm}\includegraphics[width=73mm]{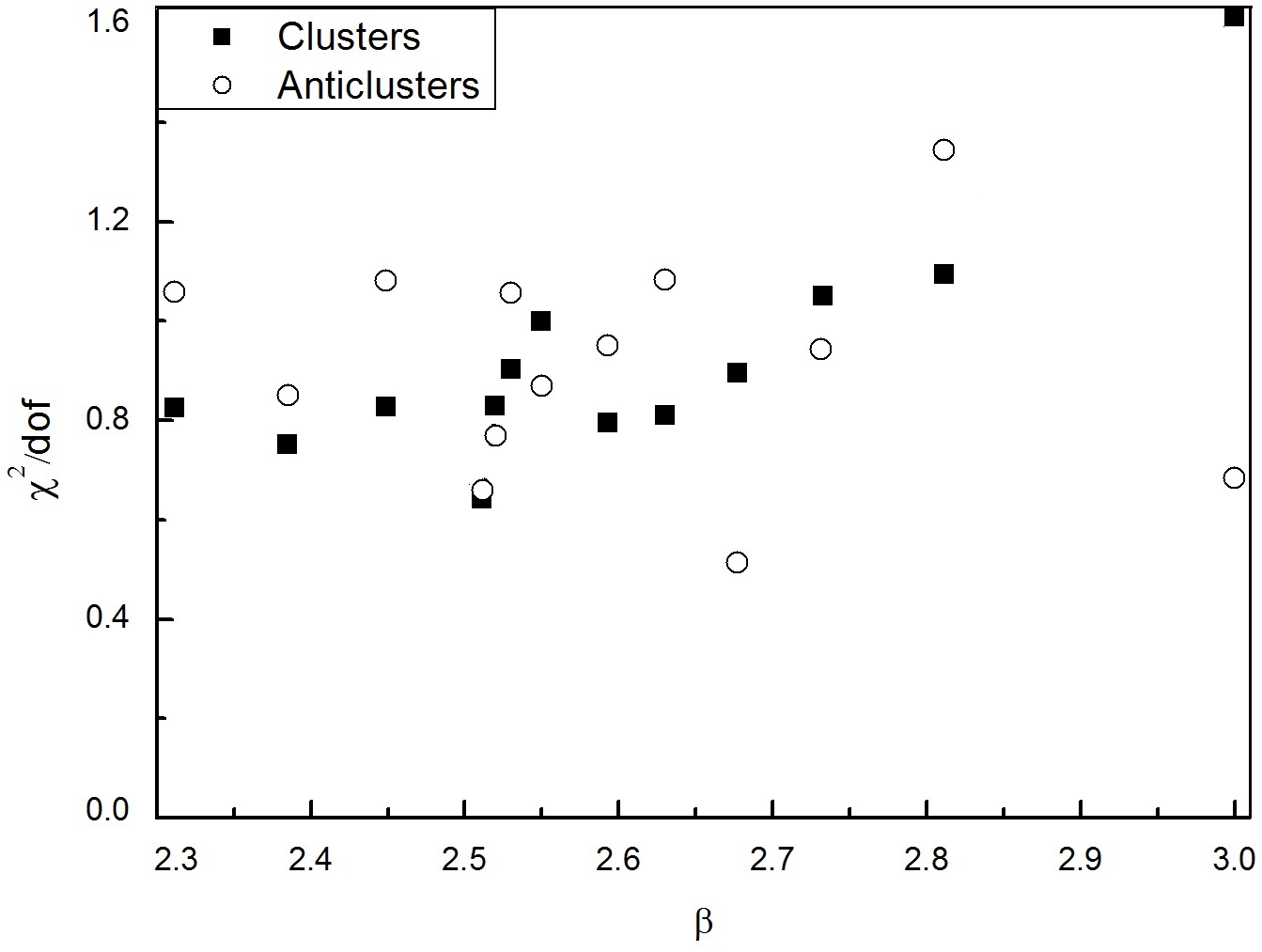}}

\caption{Results of the 3-parametric fit for the $\beta$-dependence of the 
reduced chemical potential $\mu_A$, of the
reduced surface tension $\sigma_A$, of the normalization  constant $C_A$ 
and of the mean deviation squared per degree of freedom $\chi^2/dof$. 
The fit parameters are  
shown 
for $k_{min} = 2$ at fixed $\tau_A =1.806\pm0.008$ for the data 
obtained   for  cut-off 
$L_{cut}=0.2$. In the panels depicting $\mu_A$ and $C_A$ the curves are shown 
to guide the eye, while in the panel of $\sigma_A$ the curves represent 
Eq. (\ref{EqXVII}). More details are given in the text. 
}
\label{fig6}
\end{figure}

{\bf Results of the  3-parametric fit.} From the preceding discussion it is 
clear that now we can fix $k_{min} = 2$ and
$\tau_A =1.806 \pm0.008$ and perform a fit of  (anti)cluster  distributions 
with only 3 parameters $C_A$, $\mu_A$ and $\sigma_A$  with A=\{cl, acl\}.  
These results are presented in Fig. \ref{fig6}. One can 
see that the behavior of the LDM parameters  $C_A$, $\mu_A$ and 
$\sigma_A$ for clusters and anticlusters is identical up to $\beta \le 2.5115$,
 while already 
for $\beta \ge 2.52$ the parameters of clusters and anticlusters behave 
differently. Such a difference is rooted in the symmetry breaking between 
the  cluster and anticluster distributions which occurs due to the PT at the critical 
value of $\beta$, which in an infinite system is $\beta_c^\infty=2.5115$ 
\cite{Ref_beta_crit} for the lattice with $N_\tau =8$.

Although the 3-parametric fit quality is overall good (see Fig. \ref{fig6}),  
for $\beta > 2.6$ we observe some fluctuations  
in the $\beta$-dependence of fitting parameters of anticlusters. Such a 
behavior is seen in Fig. \ref{fig6} in the deviations of 
anticluster chemical potential, surface tension and normalization constant 
from the regular curves.  On the other hand,  the behavior of the
cluster fitting parameters is perfectly regular. The main reason for such 
a difference  between clusters and anticlusters  at  $\beta > 2.6$ is that -- compared to monomers and 
dimers --  the multiplicities of  anticlusters 
with $n>3$ are extremely small and hence
they experience strong statistical   fluctuations from one lattice configuration to another. 
As a result for these values of $\beta$ a reliable statistics exists for the 
gas of anticlusters, if their volume is up 10--15 elementary lattice cubes.
This is clearly seen, if one compares the anticluster size distributions 
for $\beta = 2.53$ and $\beta = 3.0$ shown in Fig. \ref{fig2} for the 
cut-off $L_{cut}=0.2$. For the cut-off $L_{cut}=0.1$ the statistics for 
the gas of anticlusters  with size larger than 10 units  is even less reliable at $\beta > 2.6$ due to  their very 
small multiplicities.
Due to this reason in the present  
work  the main attention is paid  to the analysis of the lattice results 
obtained for the cut-off $L_{cut}=0.2$, while in appropriate places we 
comment in what respect the results obtained for the cut-off $L_{cut}=0.1$  
differ from the ones found for  $L_{cut}=0.2$.

\begin{figure}[ht]
\centerline{\includegraphics[width=73mm]{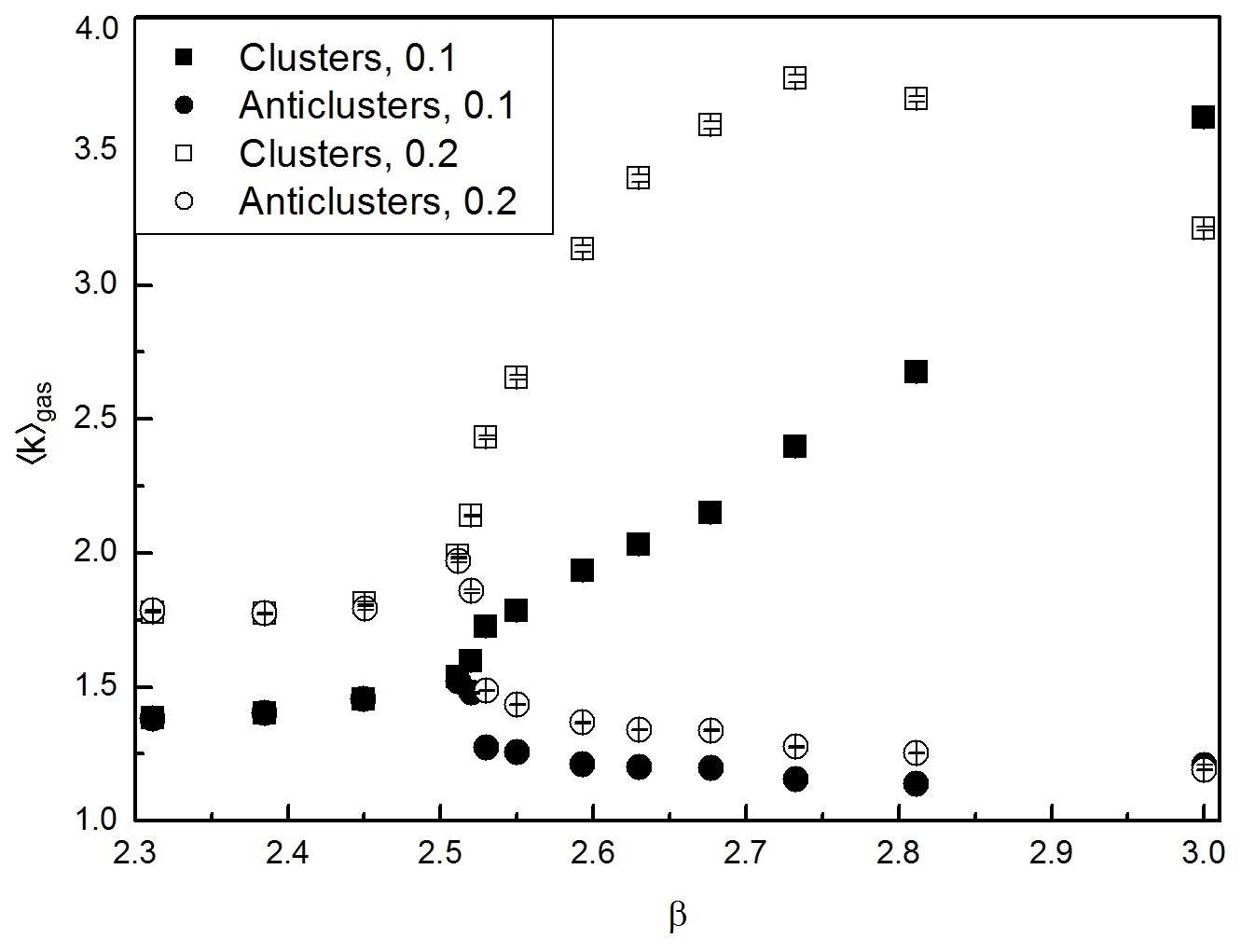} \hspace*{-1.1mm}\includegraphics[width=73mm]{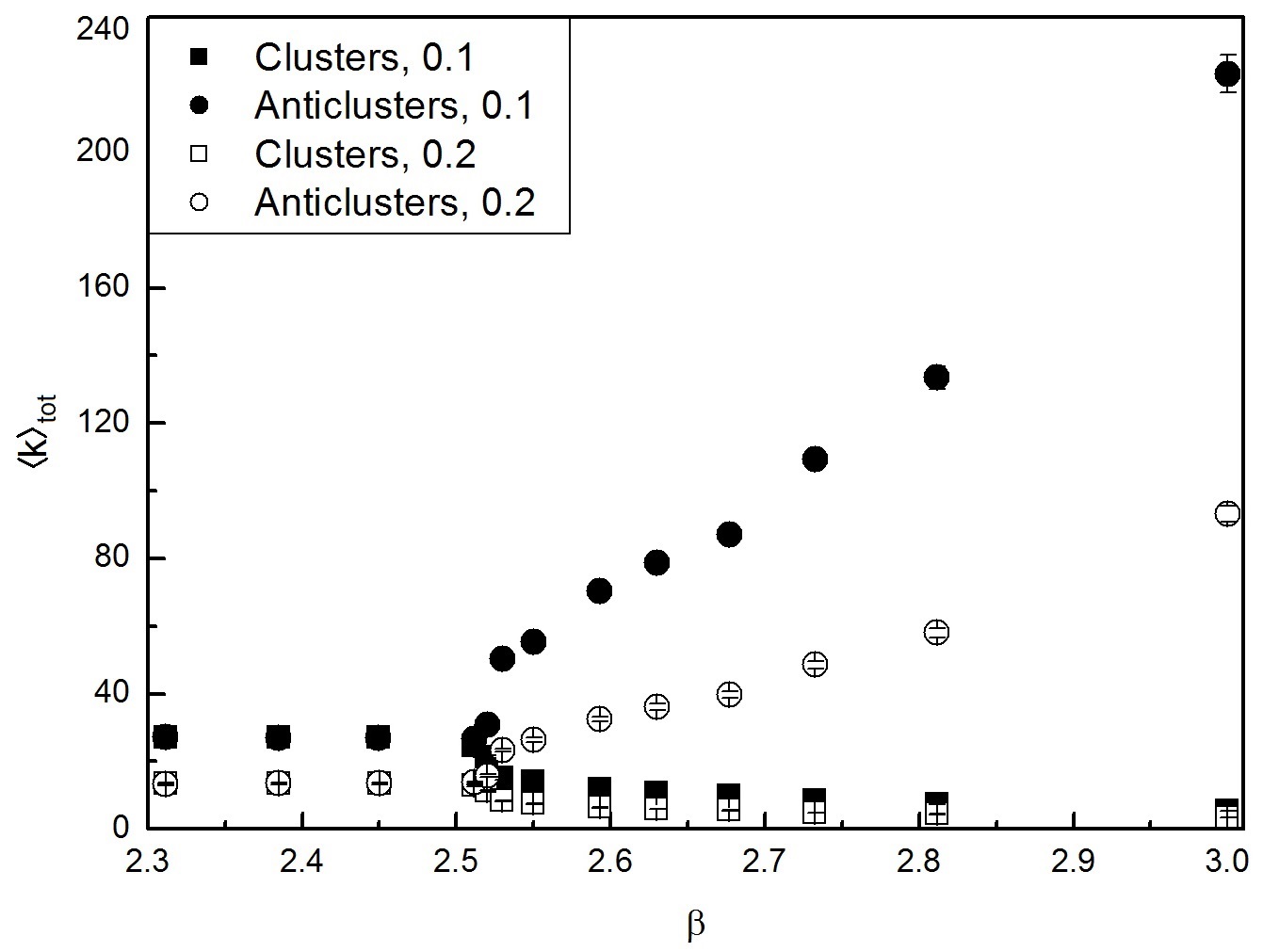}}
\caption{The $\beta$-dependence of the mean cluster volume 
$\langle k_A \rangle_{gas}$ in the A-gas (left) and the mean total 
cluster volume $\langle k_A \rangle_{tot}$ including 
the A-gas and the A-liquid droplet (right). 
Results are shown for $L_{cut} = 0.1$ and  $L_{cut} = 0.2$.}
\label{fig7}

\end{figure}

\begin{figure}[h]
\centerline{\includegraphics[width=73mm]{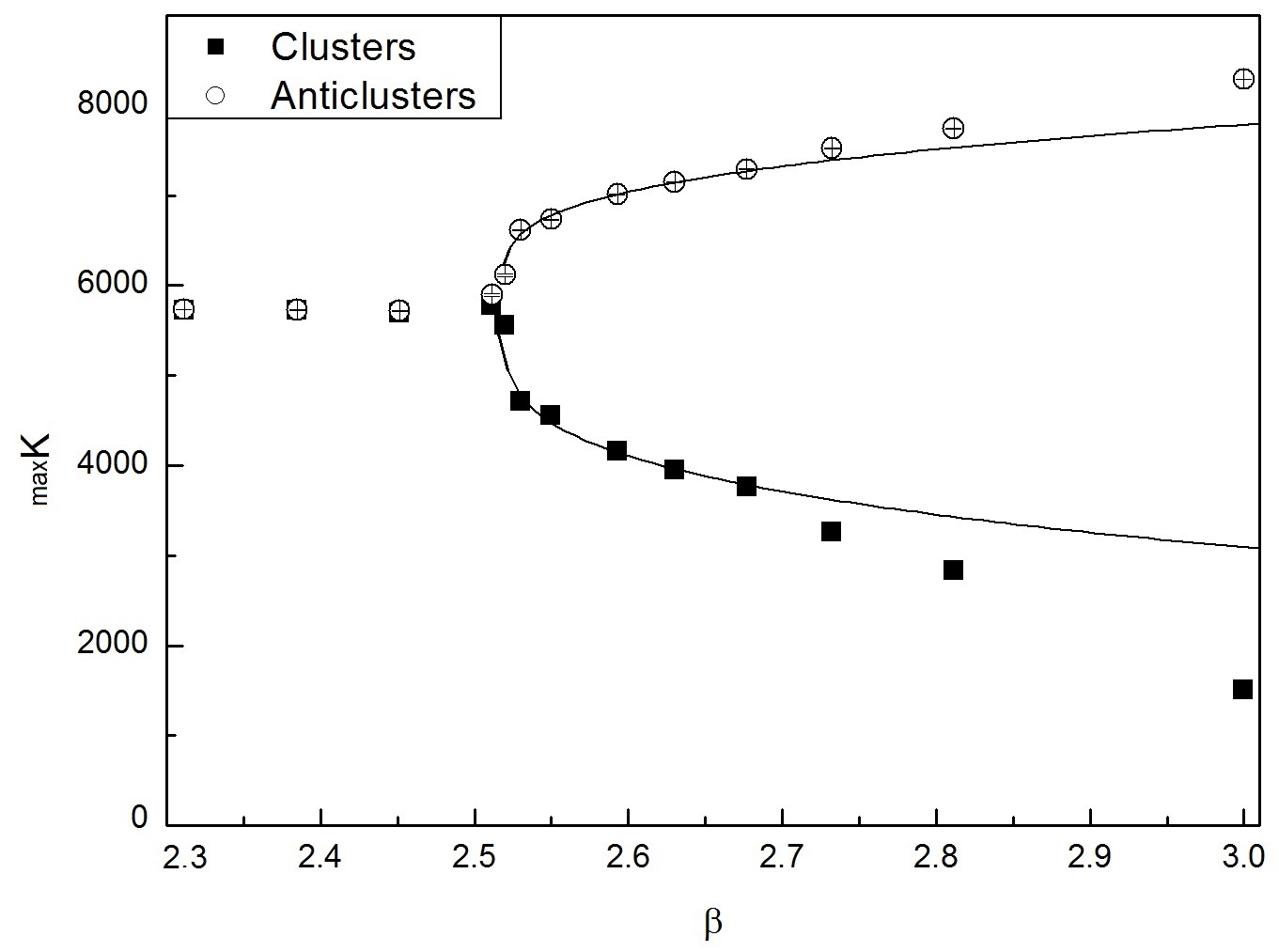} \hspace*{-1.1mm}\includegraphics[width=73mm]{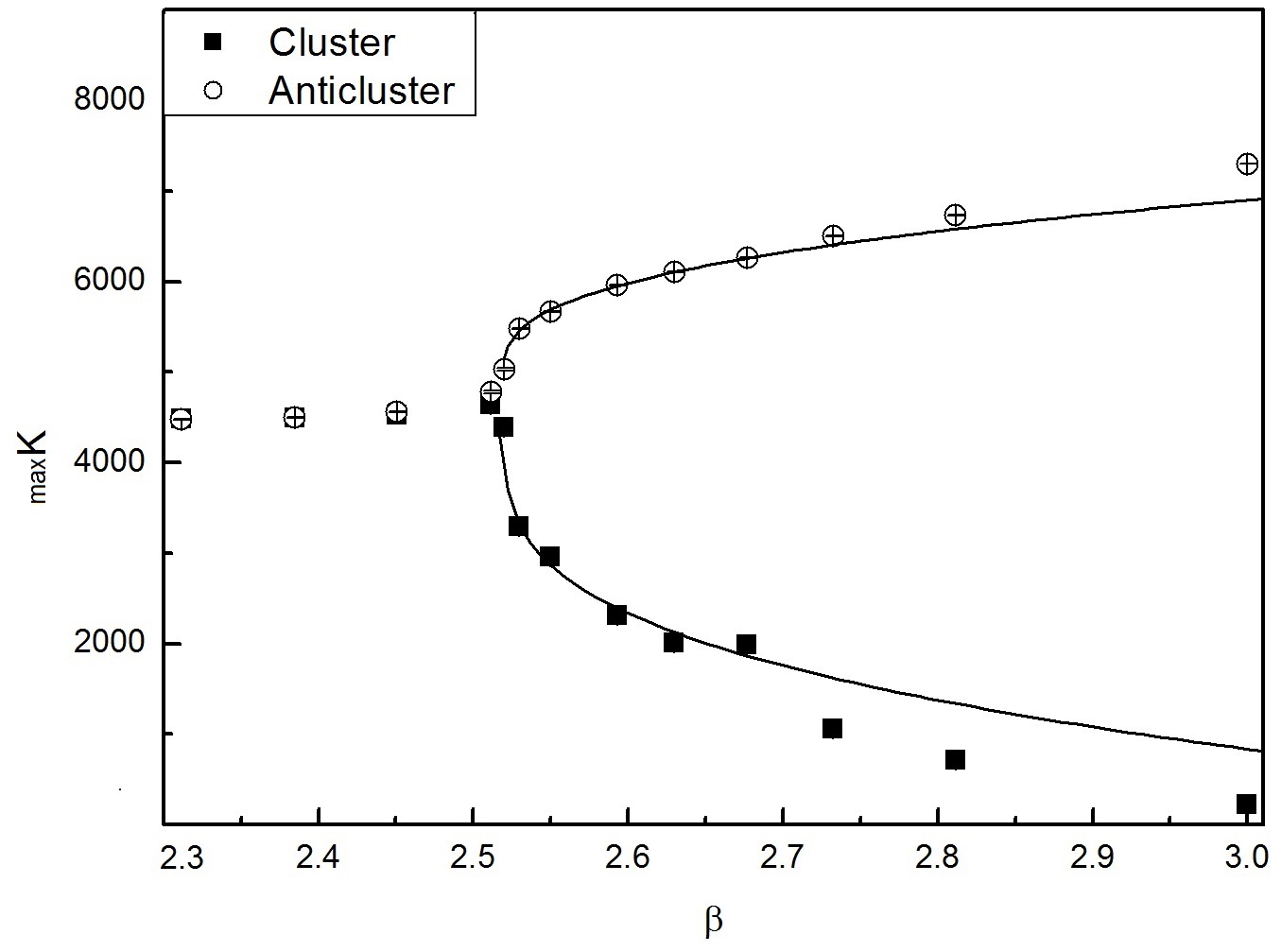}}
\caption{The $\beta$-dependence of the volume of the largest anticluster
and the largest cluster.   
Results are shown for $L_{cut} =0.1$ (left) and $L_{cut} =0.2$ (right). 
The curves represent Eq. (\ref{EqXVI}). More details are given in the text. 
  }
\label{fig8}
\end{figure}

Already from Figs. \ref{fig2}  and \ref{fig6} one can see that the real situation in SU(2) 
gluodynamics is much more complicated than just a percolation of a largest 
anticluster. 
The physical picture can be well formulated in terms of two ``substances"
which are represented by gases of clusters and anticlusters and their respective
liquids, i.e. the largest cluster and anticluster.  
The behavior of the reduced chemical potentials $\mu_A$  in  
Fig. \ref{fig6}  shows that at  $\beta \le \beta_c^\infty$ these  two ``substances'' 
 are in chemical equilibrium, while at  
$\beta > \beta_c^\infty$ it gets lost. 
This conclusion is further substantiated in 
Figs.  \ref{fig7} and \ref{fig8} which  show  the mean size of  
(anti)clusters in the respective gas, $\langle k_A \rangle_{gas}$, the 
mean size of all 
(anti)clusters $\langle k_A \rangle_{tot}$  
and the mean size of the largest (anti)clusters. 
The latter is calculated  numerically as
\begin{eqnarray}\label{EqV}
\max{K_A} = \frac{\sum\limits_{k=1} k^{1+\tau_A}\, n_A^{lat} (k)}{\sum\limits_{k=1} k^{\tau_A}\, n_A^{lat} (k)} \,,
\end{eqnarray}
where $n_A^{lat} (k)$ is the multiplicity of lattice clusters of A-type and 
volume $k$, while the value of $\tau_A$ coincides with the one found from 
the 4-parametric fit $\tau_A = 1.806 \pm0.008$. Numerically we 
have verified that Eq. (\ref{EqV}) correctly defines the mean size of the 
largest (anti)cluster for the whole range of $\beta$ values under 
investigation.

From Fig.  \ref{fig7}
one can immediately see  that  above $\beta_c^\infty$ the anticluster liquid droplet accumulates  the gas particles while the cluster liquid droplet 
evaporates them. 
Below $\beta_c^\infty$ the main contribution to both gases is given by 
monomers, dimers and trimers. Above $\beta_c^\infty$
the gas of clusters gains sizable contributions from  4-, 5- and 6-mers, 
while the gas of anticlusters quickly degenerates into a mixture of monomers 
and dimers, at most.  It is clear that the bifurcation singularity at  $\beta =\beta_c^\infty$ which is  visibly  seen in Figs. \ref{fig6}-\ref{fig8} can be identified as the critical point for the 2-nd order  PT (see further discussion in Sect. V).


Note that exactly solvable models dealing with a
1-st order  PT \cite{Bugaev_00,GasOfBags:81}, when rigorously analyzed in 
a finite volume  \cite{Bugaev:05, Bugaev:07}, 
predict that (i) in a finite system the analog of a PT must occur at positive 
values of the effective chemical potential and that (ii)
the analog of mixed phase in a finite system consists of states which are 
{\it not} in chemical equilibrium with each other. 
Both of these statements are valid also in case of a tricritical endpoint, 
at which a 1-st order PT degenerates into a 2-nd order one.  
Therefore, the predictions of 
Refs.   \cite{Bugaev:05, Bugaev:07} look very similar to what 
we find here for the behavior of the reduced chemical potentials $\mu_A$ 
of (anti)clusters (see Fig.  \ref{fig6}). 
However, there is a principal difference between the 
 solutions found in \cite{Bugaev:05, Bugaev:07}, where
the  chemical equilibrium is lost between the 
gas of all clusters (stable state) and metastable states of the same ``substance'' \cite{Bugaev:05, Bugaev:07}, 
and  the present model, where  the chemical equilibrium is apparently lost between the two
``substances'',  i.e. two gases,  as follows from the different behavior 
of their reduced chemical potentials $\mu_A$ shown in Fig.  \ref{fig6}.

\section{Discussion of physical surface tension}
\begin{figure}[ht]
\centerline{\includegraphics[width=73mm]{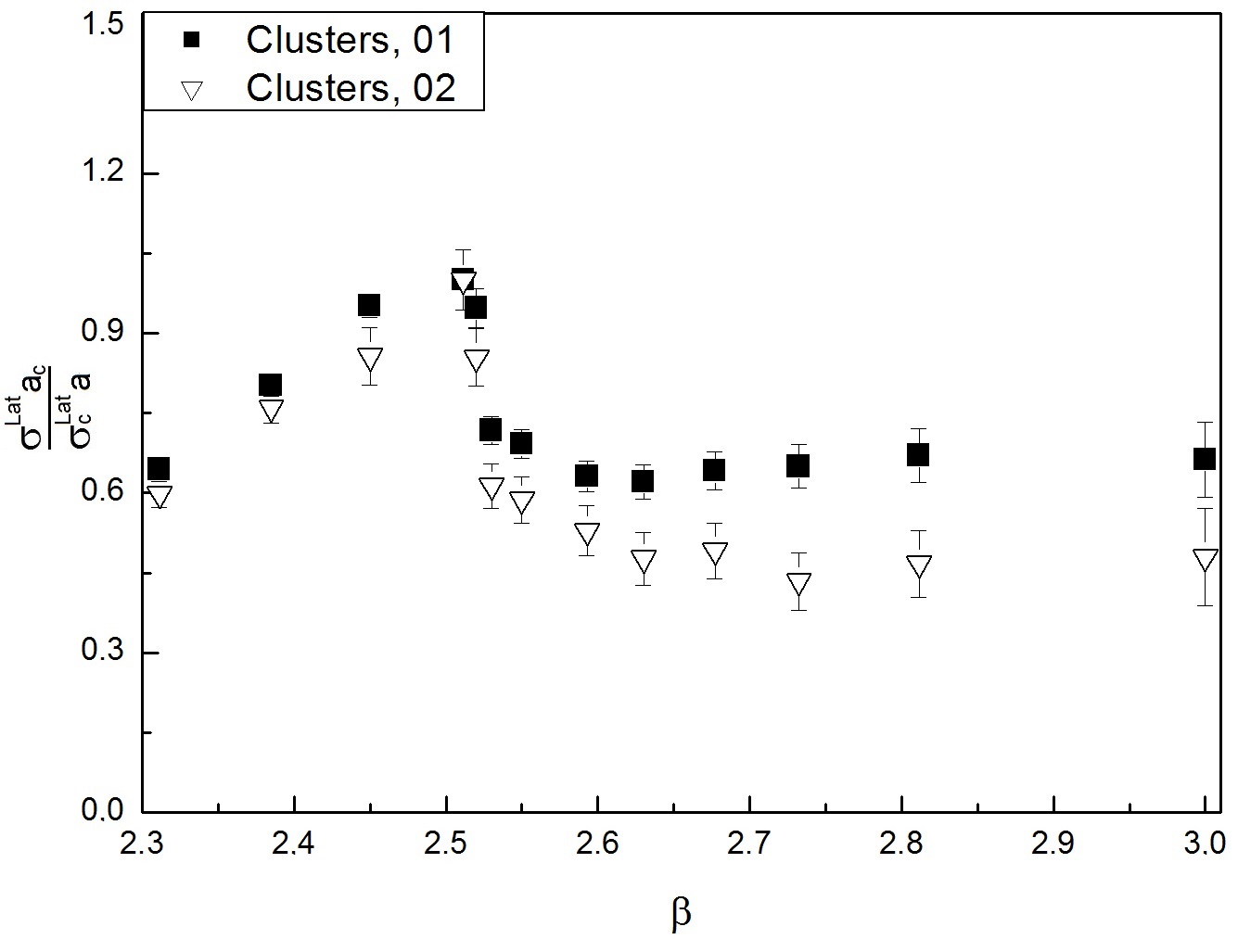} \hspace*{-1.1mm}\includegraphics[width=73mm]{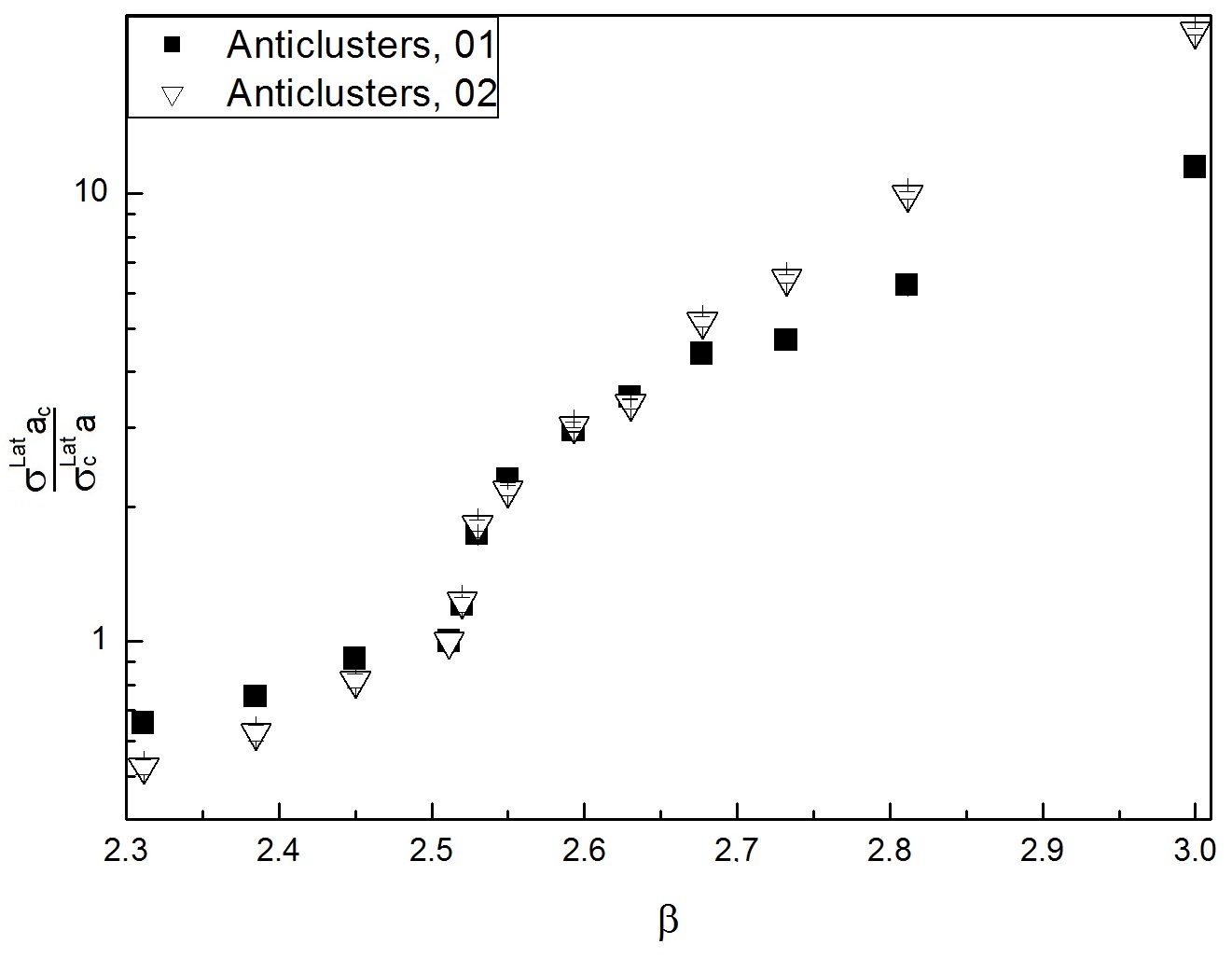}}
\caption{The $\beta$-dependence of the lattice surface tension  in relative 
units $\frac{\sigma_{A} (\beta)\, a_\sigma(\beta_c^\infty)}{\sigma_{A} (\beta _c^\infty)\, a_\sigma(\beta)}$  
 is shown 
 for clusters (left) and anticlusters (right)  for two cut-off 
values $L_{cut} = 0.1$ and  $L_{cut} = 0.2$.
 }
\label{fig9}
\end{figure}

The second principal difference between the present  and the other 
cluster models \cite{Fisher-67,Bondorf,Cluster:1,Cluster:2,Bugaev_00,Reuter_01,Bugaev_07,Bugaev_09,Ivanytskyi,Ivanytskyi2,Ivanytskyi3,sagun-14}  
is an entirely different temperature behavior of the surface tension.   
In traditional cluster models the surface tension is a decreasing function of the
temperature, and  it vanishes at the (tri)critical endpoint temperature $T_{end}$. 
At temperatures above $T_{end}$ the surface tension of existing models is either 
zero as in 
\cite{Fisher-67,Bondorf,Cluster:1,Cluster:2,Bugaev_00,Reuter_01} or it becomes 
negative as in 
\cite{Bugaev_07,Bugaev_09,Ivanytskyi,Ivanytskyi2,Ivanytskyi3,sagun-14}.
As one can see from Fig. \ref{fig6} and more clearly from Fig. \ref{fig9} the 
surface tension of (anti)clusters does not vanish at $\beta_c^\infty$ and above 
it. 
In the present analysis (with fixed $N_\tau$), the physical temperature is a 
monotonically increasing function of $\beta$ (see Table I for details) 
\cite{Beta:values} and, hence, our comparison of $\beta$-dependencies shown in 
these figures is justified.
\footnote{The values of physical temperature shown in Table I are calculated 
in a standard way using the two loop formula \cite{Beta:values}.  
Although below $\beta_c^\infty$ such an approximation can be used for a restricted 
range of values of the inverse coupling constant squared $\beta$, it is a good
approximation 
above $\beta_c^\infty$.}

\begin{table}[htp]
\caption{Two-loop $\beta$ dependence of the spatial lattice spacing 
$a_\sigma(\beta)$ given by the ratio to $a_\sigma(\beta_c^\infty)$ 
as function of the physical temperature $T$ in units of the critical 
temperature $T_c^\infty$.}
\begin{center}
\begin{tabular}{|c|c|c|}
\hline
$\beta$     &      $a_\sigma(\beta)/a_\sigma(\beta_c^\infty)$     &    $T/T_c^\infty$\\
\hline
  ~~ 2.3115 ~~ &    ~~1.7132 ~~ &   ~~ 0.5837~~ \\
   2.3850   &    1.4057   &    0.7114 \\
   2.4500   &    1.1783   &    0.8487 \\
   2.5115   &    1.0000   &    1.0000  \\
   2.5200   &    0.9774   &    1.0231  \\
   2.5300   &    0.9514   &    1.0510 \\
   2.5500   &    0.9016   &    1.1092  \\
   2.5930   &    0.8030   &    1.2453  \\
   2.6300   &    0.7269   &    1.3757  \\
   2.6770   &    0.6405   &    1.5612  \\
   2.7325   &    0.5516   &    1.8128  \\
   2.8115   &    0.4459   &    2.2423  \\
   3.0000   &    0.2685   &    3.7244  \\
\hline
\end{tabular}
\end{center}
\label{table1}
\end{table}

The  $\beta$-dependence of  physical surface tension  defined as
\begin{eqnarray}\label{EqVI}
\sigma^{phys}_A (\beta) ~\equiv ~T  \frac{ \sigma_A (\beta)}{[\, a_\sigma(\beta) \,]^2} ~= T_c^\infty ~ \frac{a_\sigma(\beta_c^\infty)}{a_\sigma(\beta)} \frac{ \sigma_A (\beta)}{[\, a_\sigma(\beta) \,]^2}\,,
\end{eqnarray}
where in the last step we have used the following relation for the temperature
$T = T_c^\infty \frac{a_\sigma(\beta_c^\infty)}{a_\sigma(\beta)}$ \cite{Beta:values}.
Such a surface tension has the correct physical dimension, but it is more convenient to use the dimensionless ratio 
${\sigma_{A} (\beta)\, a_\sigma(\beta_c^\infty)}/{\sigma_{A} (\beta _c^\infty)/a_\sigma(\beta)}$ 
because such a ratio, as one can see from Fig. \ref{fig9}, clearly demonstrates 
the 
different behavior on two sides of the point  $\beta=\beta _c^\infty$. 
Also in this case one does not need to care about an exact value of a coefficient 
relating the volume and the surface of (anti)clusters.   
To find this ratio  from Eq. (\ref{EqVI}) we used the second column of Table I.

We would like to stress that in contrast to all known cluster models the physical 
surface tension of clusters has a peak at about 
$\beta _c^\infty$ while the surface tension of anticlusters has a kink at this 
point.
Also  both of them 
show a jump of the 
derivative with respect to $\beta$ and, hence, a jump of the
derivative with respect to $T$, on two  sides of the point
$\beta =\beta _c^\infty$. Such a behavior of the surface tension  
would be rather unusual for the 1-st order liquid-gas PT in traditional  
cluster models 
\cite{Fisher-67,Bondorf,Cluster:1,Cluster:2,Bugaev_00,Reuter_01,Ivanytskyi}, 
whereas it was recently suggested in \cite{Bugaev_09} as a mechanism of a 
surface-tension induced deconfining transition in a phenomenological model 
constructed with the aim to explain a PT in QCD with a critical endpoint. 
However, in contrast to the present findings, the deconfined phase in 
the  model of  \cite{Bugaev_09} has a negative surface tension. 
Also it is necessary to 
remember that for most of existing cluster models with a 1-st order PT of 
liquid-gas type there is no jump of $T$-derivative of surface tension on 
two sides of the (tri)critical endpoint temperature $T_{end}$, 
where  the 1-st order PT changes into a 2-nd order PT.
There are two well established exceptions presented by models which have a critical
endpoint,
namely the Fisher droplet model \cite{Fisher-67} and the QGBST 
model \cite{Bugaev_09, Ivanytskyi,Ivanytskyi2}. 
But
both  these models are valid for the Fisher exponent value $\tau > 2$, 
while the present model, derived from SU(2) gluodynamics, has a clearly different
Fisher exponent $\tau_A \simeq 1.806 < 2$.

{The fact that in the vicinity of critical temperature  $T = T_c^\infty$ the physical surface tension 
of (anti)clusters does not vanish is in contrast with  the traditional belief  of the cluster models that 
at the critical or  tricritical endpoint there should exist the power law in the size distribution
of  physical clusters.  In the present study we see that the anticlusters do not show any power law at any $\beta$,
while the  power law of the size distribution of clusters is seen at $\beta =3.0$, i.e. at very  high
temperature.
}

This leads us to the first important conclusion of the present work. 
Our analysis of the SU(2) Polyakov loop clusters and anticlusters 
demonstrates that none of existing cluster models of the liquid-gas PT can be 
used to safely interpret the results. Our discussion above shows that the gases 
of clusters and anticlusters have a few common features with some of the models, 
but their main properties such as the surface tensions and the apparent absence of 
chemical equilibrium above $T_c^\infty$ cannot be described by existing models. 
Therefore, the new properties of the gases of clusters and anticlusters in 
SU(2) gluodynamics require a development of a new 
cluster model which differ from the existing ones.

\begin{figure}[h]
\centerline{\includegraphics[width=73mm]{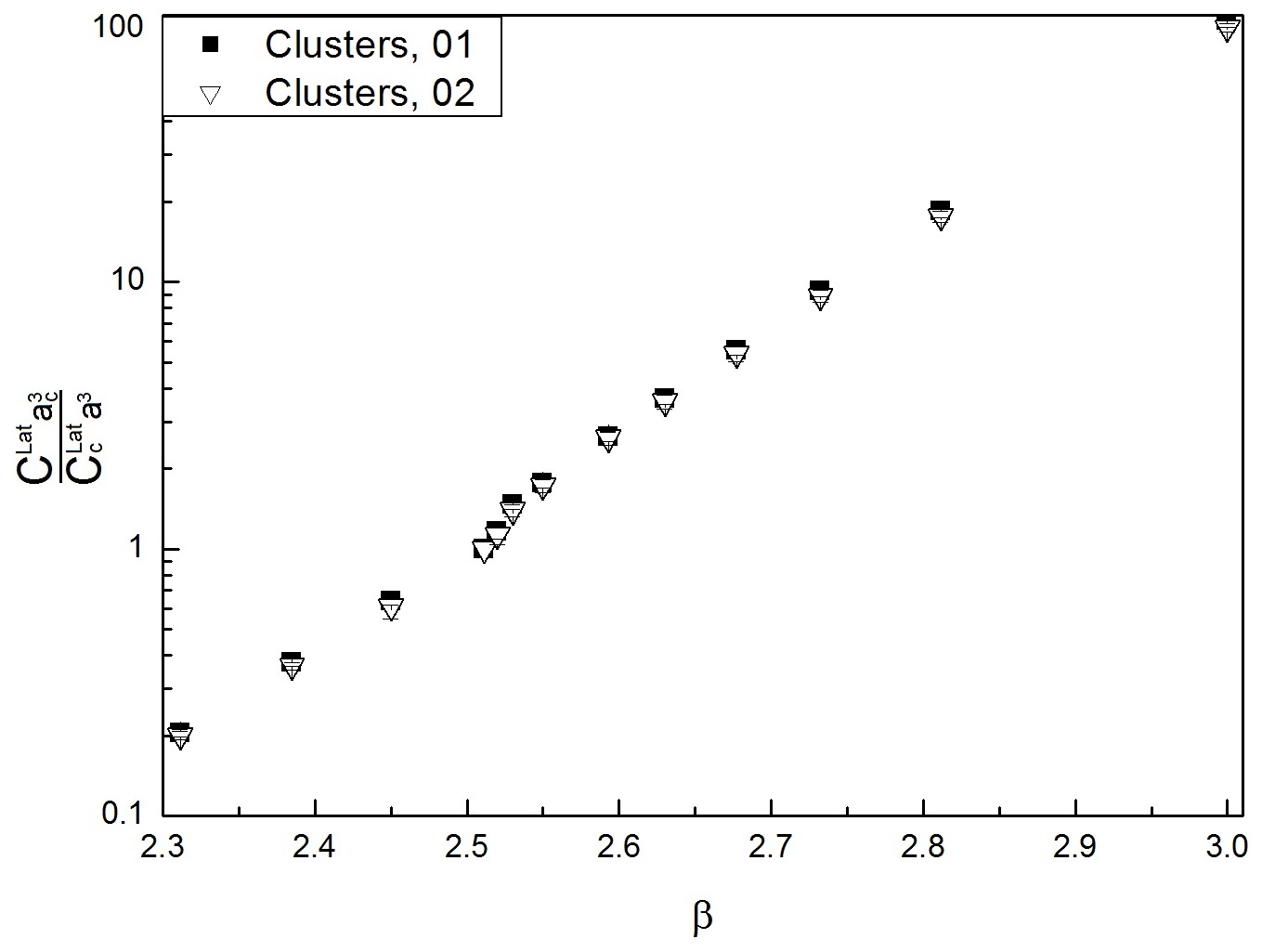} \hspace*{-1.1mm}\includegraphics[width=73mm]{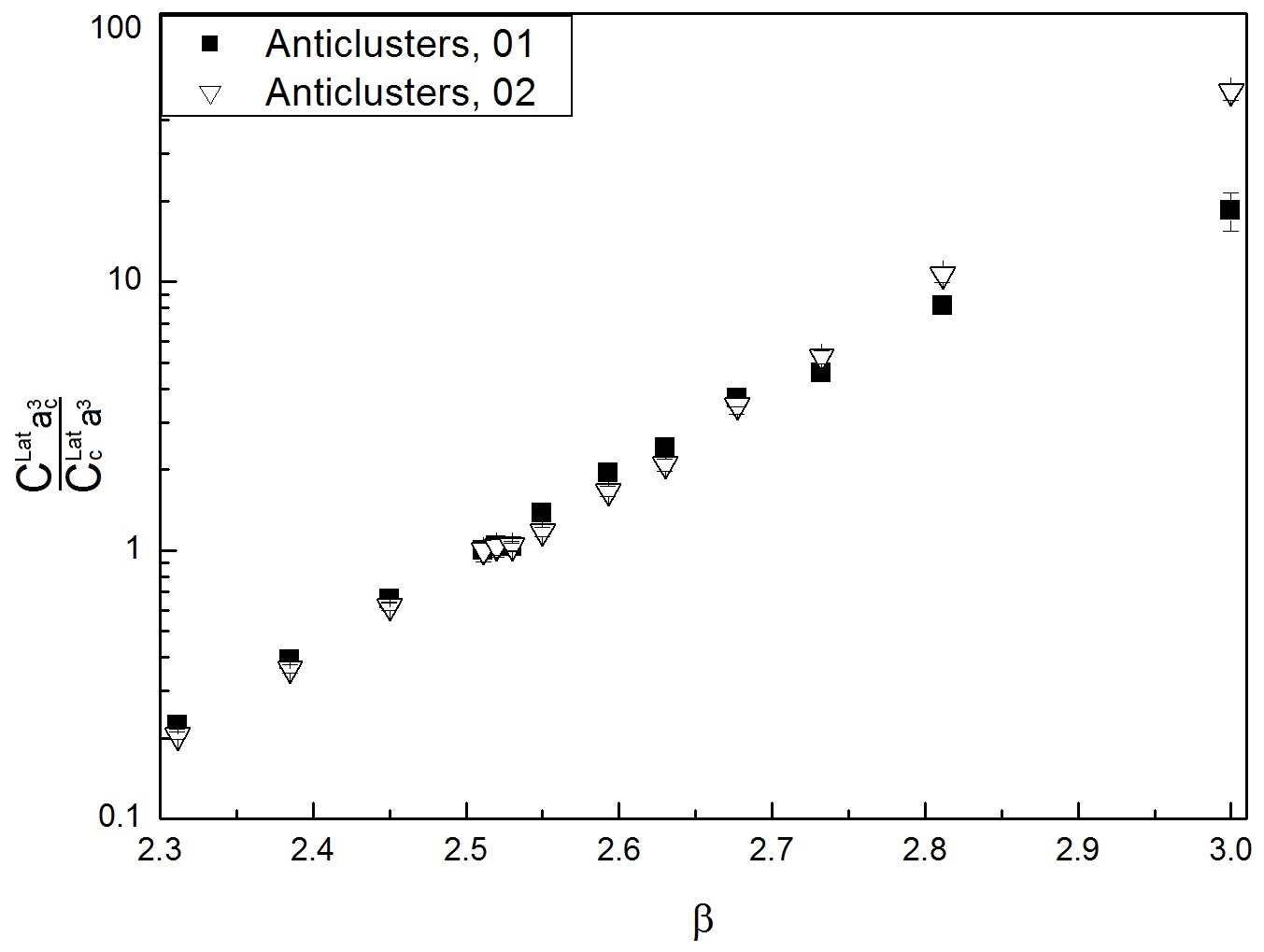}}
\caption{Same as in Fig. \ref{fig9}, but for the normalization factors  in physical units $ \frac{C^{phys}_A (\beta)}{ C^{phys}_A(\beta_c^\infty)}$ according to Eq. (\ref{EqVII}).
  }
\label{fig10}
\end{figure}

From Fig. \ref{fig9} one can also see that the behavior of the normalized surface 
tension of (anti)clusters is rather similar for both values of the cut-off parameter.
Of course, there is no absolutely identical behavior, but, first of all, this is 
caused by essentially smaller statistics  of data studied with  $L_{cut}=0.1$ 
(especially for anticlusters). Secondly,  all quantities 
referring to clusters and anticlusters, if normalized to their values at  
$\beta _c^\infty$,  demonstrate a very similar behavior  as function of $\beta$ for both cut-offs.
Apparently, the list of such quantities  includes not only $\mu_A$, $\sigma_A$ 
and $C_A$, but also the average volume of (anti)clusters in the gas 
phase, $\langle k_A \rangle_{gas}$ (see Fig. \ref{fig7}), as well as
their total average volume  $\langle k_A \rangle_{tot}$ (see Fig. \ref{fig7}) 
including the largest clusters of both kinds,  as well as the mean volume
of the largest clusters of both kinds defined by Eq. (\ref{EqV}).
To illustrate the convenience of normalized quantities once more, 
in Fig. \ref{fig10} we show the behavior of properly scaled  normalization factor $C_A$ 
\begin{eqnarray}\label{EqVII}
 \frac{C^{phys}_A (\beta)}{ C^{phys}_A(\beta_c^\infty)} ~\equiv ~   \frac{C_{A} (\beta)}{ C_{A}(\beta_c^\infty)}  \left[ \frac{a_\sigma(\beta_c^\infty)}{a_\sigma(\beta)} \right]^3  \,.
\end{eqnarray}
Such a choice is made  in order to get the particle number density of  $k$-volume (anti)clusters $\rho_A(k)$. To find it  one has to
divide the multiplicity of $k$-volume (anti)clusters, $n_A^{lat}(k)$, observed  on the lattice  by the 
system volume 
$V = \left[ N_\sigma \, a_\sigma(\beta) \right]^3$. In this case the normalization 
factor $C_{A} (\beta)$ in physical units acquires 
the form (\ref{EqVII}). Fig. \ref{fig10} demonstrates that in physical units the 
normalization factors of clusters which are found for two cut-off values are 
practically identical, while for anticlusters they slightly differ  
at high values of $\beta$ at which the statistics for a cut-off $L_{cut} =0.1$ 
gets rather poor.  
Also this figure shows that in physical units the kink in the $\beta$ dependence  
of normalization factor $C^{phys}_A$ is hardly noticeable.

Note, that according to the definition (\ref{EqII}) the normalization factor 
$C^{phys}_A$ in physical units $C^{phys}_A$ has a meaning of thermal density 
of  $k$-volume (anti)clusters per one internal degree of freedom 
(i.e. the total particle number density per one internal degree of freedom at $\mu_A =0$), 
while the factors depending on 
$\sigma_A$ and $\tau$ in (\ref{EqII})  determine the number of  internal degrees of 
freedom \cite{Fisher-67}.
As one can see from Fig. \ref{fig6}, the normalization factor of anticlusters 
$C_{acl}$ already saturates 
at $\beta = 2.7$, while such a quantity for clusters $C_{cl}$ shows a weak 
linear growth for $\beta \ge  2.7$. However, by changing $\beta$ from 2.81 
to 3.0 we increase $C_{cl}$  by 10\%, while the physical temperature is 
increased by 66\%. Thus, there is a very weak linear temperature dependence 
of $C_{cl}$ at  $T/T_c^\infty \ge  2.24$, which  can be safely ignored.
{One can make a rough estimate  of the temperature dependence power for the clusters noting that  the ratio $ \frac{C^{phys}_{cl} (\beta)}{ C^{phys}_{cl}(\beta_c^\infty)} $ changes in 500 times (see left panel of  Fig. \ref{fig10}), while temperature increases in about 6.3 times. Then the desired power
for clusters is $\frac{\ln(500)}{\ln(6.3)} \simeq 3.3$. For the anticlusters one can estimate this power by considering a change of the ratio 
 $ \frac{C^{phys}_{acl} (\beta)}{ C^{phys}_{acl}(\beta_c^\infty)} $ within the interval $\beta \in [2.3115, 2.8115]$ at which the temperatures 
 increases in about 3.84 times. From the right panel of  Fig. \ref{fig10} one finds that this power  for anticlusters is $\frac{\ln(50)}{\ln(3.84)} \simeq 2.9$.  

From these estimates  we conclude that at high temperatures, i.e. 
for $1.25\, T_c^\infty < T  \le 3.7 \, T_c^\infty$, 
the normalization factor $C^{phys}_A$ in physical units behaves similarly 
to the particle  number  density  in ultrarelativistic gas  (a gas of massless particles), 
i.e. as $C^{phys}_A \sim T^3$. Note that the only massless particles which exist in gluodynamics are the gluons, whose thermal density behaves as $T^3$ \cite{Karsch:2003jg},  although the gluons  were not 
 explicitly considered in our study.
Thus, we believe it is remarkable that information about 
massless gluons is implicitly encoded in the density  of  
Polyakov loop (anti)clusters.
}

Now we would like to discuss some  physical consequences of the temperature 
dependence of the surface tension of clusters and anticlusters.
Although in the present work we do not investigate the thermodynamic limit of 
the LDM parameterization,  we can  draw certain conclusions on the temperature 
dependence of cluster properties. The importance of the surface tension of quark-gluon (QG) bags was 
realized a long time ago \cite{Jaffe:1,Svet:1,Madsen}. However, despite of many 
efforts to calculate this quantity in the modern literature,  the predicted surface tension of QG bags at zero 
temperature ranges from 5-15 MeV/fm$^2$ \cite{SurfT:s1,SurfT:s2} to about 
150 MeV/fm$^2$ \cite{SurfT:m1,SurfT:m2} or even 300 MeV/fm$^2$ \cite{SurfT:l1}.  
Also it is not much known about the QG bag surface tension behavior at high 
temperatures \cite{SurfT:m1,SurfT:m2}.  
Without studying the thermodynamic limit of the present approach, we cannot  
fix an absolute magnitude  of the coefficient in Eq. (\ref{EqVI}) for the 
physical surface tension. However, the present approach allows us  to predict 
the functional $T$-dependence of the physical surface tension of Polyakov loop clusters
\begin{eqnarray}
\sigma^{phys}_{cl} (T)~=~  \frac{Const}{[\, a_\sigma(\beta) \,]^2} ~\sim ~T^2 \quad {\rm for} \quad 1.25\, T_c^\infty < T   \le  3.7\, T_c^\infty\,.
\end{eqnarray}
The validity of this estimate is clearly seen from Fig. \ref{fig9}, which 
demonstrates that the ratio  ${\sigma_{cl} (\beta)\, a_\sigma(\beta_c^\infty)}/{\sigma_{cl} (\beta _c^\infty)/a_\sigma(\beta)}$ 
saturates at $\beta \ge 2.6$. Also we found the following estimate 
$\sigma^{phys}_{acl} (T)~  \sim ~T^4$ for the physical surface tension 
  of anticlusters 
at temperatures belonging to the 
range $1.25\, T_c^\infty < T   \le  3.7 \, T_c^\infty$. However, 
one should keep in mind that, in contrast to clusters, for which   $\mu_{cl} \rightarrow 0$ in this 
temperature range,  the quantity  
$\sigma^{phys}_{acl} (T)$  may also include an unknown dependence on reduced 
chemical potential $\mu_{acl}$, which for anticlusters  strongly increases 
with $T$ (see Fig.  \ref{fig6}).
 
 {One important difference between  the present consideration and the  traditional cluster models is
 that  the largest Polyakov loop  cluster (anticluster) is not homogeneous inside and it looks like a Swiss Cheese,
 since it is filled by the gas of anticlusters (clusters).  
 A similar conclusion for the largest anticluster was recently suggested in \cite{Gattr_fresh}.
 Note, however,  that in traditional cluster models, including the Fisher droplet model \cite{Fisher-67}, such a possibility  is usually ignored.  Let us demonstrate this important new feature for  the cut-off $L_{cut} =0.2$ and $\beta =3.0$.
 At  high values of $\beta$ the treatment gets simpler, since the gas of anticlusters is practically absent
 and the largest cluster is rather small. Indeed, for $\beta =3.0$ one finds that  the volume of largest anticluster is $\max K_{acl} = 7300$, the volume of largest cluster is $\max K_{cl} = 223$, the total volume of the gas of anticlusters is only
 $V^{gas}_{acl} = \widetilde{ \sum_k} \,  k\, n_{acl} (k) = 89$,  while  the total volume of the gas of clusters is $V^{gas}_{cl} = \widetilde{ \sum_k} \,  k\, n_{cl} (k) = 2848$ and the volume of auxiliary vacuum is $V_{vac} = 1707$. Hereafter  the sums with tilde indicate that the summation does not include the largest (anti)cluster.  In order to find out where the gaseous  clusters are located,  let us first estimate  the number of nearest neighbors for the gas of clusters. Since  in the gas of clusters the number of monomers  is $n_{cl} (1) \simeq 550$,  the number of  dimers is $n_{cl} (2) \simeq 130$ and the number of 
 trimers and fourmers are, respectively, $n_{cl} (3) \simeq 60$ and $n_{cl} (4) \simeq 35$, then one can estimate the number of their nearest neighbors as $N^{near}_{cl} \simeq 6 \,n_{cl} (1) + 10 \,(n_{cl} (2) + n_{cl} (3) + n_{cl} (4) ) \simeq  5600$. Here we have taken into account that each monomer has 6 nearest neighbors, each dimer has 10 nearest neighbors, while the larger clusters have at least  10 nearest neighbors.  According to definition,   the nearest neighbors of a gaseous cluster cannot be the other gaseous clusters themselves  or the largest cluster, but should be only the anticlusters or vacuum. However, the gas of anticlusters is practically absent at $\beta=3.0$, while the vacuum can provide maximum 1707 nearest neighbors out of  about 5600. Therefore, the gas of clusters must be located inside the largest anticluster, since it is the only  possibility to locate them. 

Evidently,  the number of nearest neighbors given above is somewhat overestimated. 
One can find it differently. 
Suppose that there is a  dense packing of gaseous clusters among  the vacuum and  anticlusters. The dense packing for monomers, evidently, means  that they are surrounded by the monomers of vacuum or/and  anticlusters, i.e. the volume of surrounded  matter $N^{surr}_1$ in this case equals to  the number of cluster monomer $N^{surr}_1 = n_{cl} (1) \simeq 550$.  The volume of   matter which surrounds all larger clusters can be found from the expression 
\begin{eqnarray}\label{EqXN1}
N^{surr}_2  \simeq  \sum\limits_2^{\max K_{cl}}  \, \frac{S(k)}{2} \, n_{cl} (k)  \simeq 2550\, ,
\end{eqnarray}
where $\max K_{cl} =223$ is the size of largest cluster at $\beta=3.0$,  $n_{cl} (k)$ is the number of clusters of volume $k$,
$S(k) = c_s\, k^\frac{2}{3}$ is the surface of a cluster of volume $k$, while the coefficient $c_s \simeq 4.83$ relates the surface of sphere to its volume. Note that  such a treatment is justified by the  validity of  the LDM formula. It is also  evident that  considering all clusters with the volume $k \ge 2$ as spheres we can estimate $N^{surr}_2$ from below, since  for a given volume a spherical form provides a minimal surface.   In (\ref{EqXN1}) the coefficient $\frac{1}{2}$ accounts for the fact that  for dense packing  only a half of a surrounding volume ``belongs" to a given cluster, while the other half  ``belongs" to the other cluster. In other words, this coefficient allows one to avoid the double counting of the same  surrounding volume  separating two neighboring clusters. 

Summing up $N^{surr}_1$ and $N^{surr}_2$ one can find the total surrounding volume for all clusters 
$N^{surr}_{tot} \simeq 3100$.  Even in this extreme case  the vacuum  and small gaseous anticlusters can together surround  only  1707+89 = 1796  units of  volume of gaseous clusters, while the remaining number  of gas clusters should be located inside the largest anticluster, i.e. the latter  should resemble the Swiss Cheese. 
Hence, we come to the same conclusion again. 
Note that in this case the relative packing fraction of gaseous clusters  inside the largest anticluster is 
$\rho_{cl/acl} = V_{cl}^{gas}/ \max K_{acl} \simeq 0.39$, i.e.  our previous assumption about the dense packing  was too strong. 
 
 It is evident that the gas of clusters did not exclusively  appear inside the largest anticluster at $\beta =3.0$, but this gas  existed inside of it at all temperatures. Similarly, the gas  of anticlusters exists inside the largest cluster at all values of $\beta$. 
 These conclusions naturally explain that by increasing $\beta$ above $\beta_c$ the volume of  gas of clusters  increases simultaneously with the growth of $\max K_{acl}$ and that in this case the volume of  gas of anticlusters  decreases
simultaneously with the reduction of $\max K_{cl}$.  For  $\beta =3.0$ this statement  can be nicely illustrated using the following estimates.  Indeed,  comparing the  relative packing fraction  of  the gaseous anticlusters inside the largest cluster $\rho_{acl/cl} = V^{gas}_{acl}/ \max K_{cl} \simeq 0.399$ and the one of the gaseous clusters inside the largest anticluster   $\rho_{cl/acl}  \simeq 0.39$, one finds almost the same values.  In addition, an existence of the Swiss Cheese structure would naturally  explain the fact of the fractal dimension of the largest (anti)clusters \cite{Gatt:I, Fractals}.  
Therefore,  one should distinguish between the volume of the largest (anti)cluster and its  geometrical size, which can be essentially larger, than its volume due to presence of the  non-native  gas.  Hence, 
the surface free energy of the largest (anti)cluster should have the following  form
\begin{eqnarray}\label{EqXa}
F_A^{surf} =  \Sigma_{A}^{outer} \left[ \max K_A + \widetilde{ \sum_k} \,  k\, n_{\bar A} (k) \right]^\frac{D_A-1}{D_A} 
- T \sigma_{\bar A} \widetilde{ \sum_k} \,  k^\frac{2}{3}\, n_{\bar A} (k)    
\,,
\end{eqnarray}
where the notation $\bar A$ means the summation over the  gaseous clusters which are non-native for the largest  cluster  of  
sort $A$
and $D_A$ is its fractal dimension \cite{Gattr_fresh}.
The first term on the right hand side accounts for the outer surface of 
maximal (anti)cluster containing the non-native gas with the outer surface tension coefficient $\Sigma_{A}^{outer}$,
while the second term on the right hand side of (\ref{EqXa}) accounts for the surface free energy of  all cavities made by the non-native clusters. 
Now it is also clear that  non-native gas may exist 
 inside the  (anti)clusters which are smaller than the largest one. 
 
 Another important observation related to the surface free energy of the largest anticluster is that at large values of $\beta$
it becomes negative. This is easy to understand, if one recalls that  at such values of  $\beta$ the largest anticluster  practically occupies the whole lattice. Since in all  spatial  directions the lattice fields have the periodic boundary conditions,  the  surface 
of largest  anticluster  has no outer borders (since it has the borders with itself only). Due to this reason the first term on the right hand side of Eq. (\ref{EqXa}) is absent at  large values of $\beta$ (or at high temperatures in our  formulation)  and, therefore,  its surface free energy becomes  negative, i.e. $F_{acl}^{surf}< 0$.  
Note that appearance of negative surface free energy of  largest anticluster
might be  important  for  understanding the reason of  cross-over appearance  in full  QCD at low baryonic densities  \cite{Bugaev_07,Bugaev_09, Greiner07, Greiner08}. 
 }

\section{Surface tension coefficient as a new order parameter}

From  Figs. \ref{fig6}-\ref{fig9} one can immediately learn
that in the phase of unbroken $Z(2)$ symmetry, i.e. for 
$\beta \le \beta _c^\infty$,  the behavior of   the thermodynamical functions
$\mu_A$, $\sigma_A$,  $C_A$, $\langle k_A \rangle_{gas}$,  $\langle k_A \rangle_{tot}$ and $\max{K_A}$ 
for clusters and anticlusters is absolutely identical within the error bars, 
while in the phase of broken 
$Z(2)$ symmetry these functions are  entirely different. Therefore, one 
can use the following combination 
\begin{eqnarray}\label{EqIX}
P_q = \left| \frac{q_{acl} - q_{cl}}{q_{acl}+q_{cl}} \right| \,,
\end{eqnarray}
for any of 
the quantities $q_A \in \{\mu_A, \sigma_A, C_A, \langle k_A \rangle_{gas},  
\langle k_A \rangle_{tot},  \max{K_A}\}$ as an order parameter of  PT between 
the phases with unbroken and broken symmetry.  We, however, would like to show 
that from the present results obtained for finite systems one can extract a 
more detailed physical information. 

In order to show this, first we have to discuss the traditional order parameter, 
i.e. the spatial average of  the local  Polyakov loop $\langle L(\vec x) \rangle$. 
Evidently, one can identically rewrite the average over all spatial points as 
an average over all clusters and anticlusters 
\begin{eqnarray}\label{EqX}
&&\hspace*{-8.8mm}\langle L (\vec x)   \rangle  \equiv   \frac{\sum\limits_{x} L_{x}}{N_\sigma^3}=
\frac{\sum\limits_{acl} |L_k^{acl}| \,k \, n_{acl} (k)- \sum\limits_{cl} |L_k^{cl}| \,k \, n_{cl} (k)}{N_\sigma^3}  = \nonumber ~\\
&&\hspace*{-8.8mm}\frac{\widetilde{\sum\limits_{k}} \left[ |L_k^{acl}| \,k \, n_{acl} (k)-  |L_k^{cl}| \,k \, n_{cl} (k)  \right] + |L_{max}^{acl}| \max K_{acl} - |L_{max}^{cl}| \max K_{cl}}{N_\sigma^3}  \,, \quad 
\end{eqnarray}
where
$|L_k^{cl}|$  and $|L_k^{acl}|$ denote, respectively, the modulus of  mean value of Polyakov 
loop inside  clusters and  anticlusters of volume $k$.  In the last equality we 
used the sum over the volumes of all (anti)clusters except for the largest 
one, which are added  explicitly. In doing so we accounted for the fact, that the 
largest (anti)cluster is always a single one.  
For definiteness, in Eq. (\ref{EqX}) we defined $\langle L (\vec x) \rangle$ 
in such a way that in thermodynamic limit it vanishes at and below 
$\beta_c^\infty$, but is positive above $\beta_c^\infty$.  
Eq. (\ref{EqX}) shows that vanishing of  $\langle L (\vec x) \rangle$ in the 
symmetric phase is provided by the set of inequalities 
\begin{eqnarray}\label{EqXI}
 |L_k^{acl}| + |L_k^{cl}| & \gg &  \left| |L_k^{acl}| - |L_k^{cl}|  \right|  \, ,  \\
 \label{EqXII}
  n_{acl} (k) + n_{cl} (k) & \gg & \left|  n_{cl} (k) - n_{acl} (k)  \right|   \,,
\end{eqnarray}
which are valid for all volumes $k \ge 1$. Note that in finite systems, which 
we study here using the lattice data with limited statistics,  the right hand sides 
of Eqs. (\ref{EqXI}) and (\ref{EqXII}) do not always vanish for $k \gg 10$, 
but the multiplicity of such (anti)clusters is so small compared to monomers 
or dimers (about 4-6 order of magnitude as one can see from Fig. \ref{fig2}) 
that one can safely ignore them in the sums marked with tilde in Eq. (\ref{EqX}). 

\begin{figure}[h]
\centerline{\includegraphics[width=110mm]{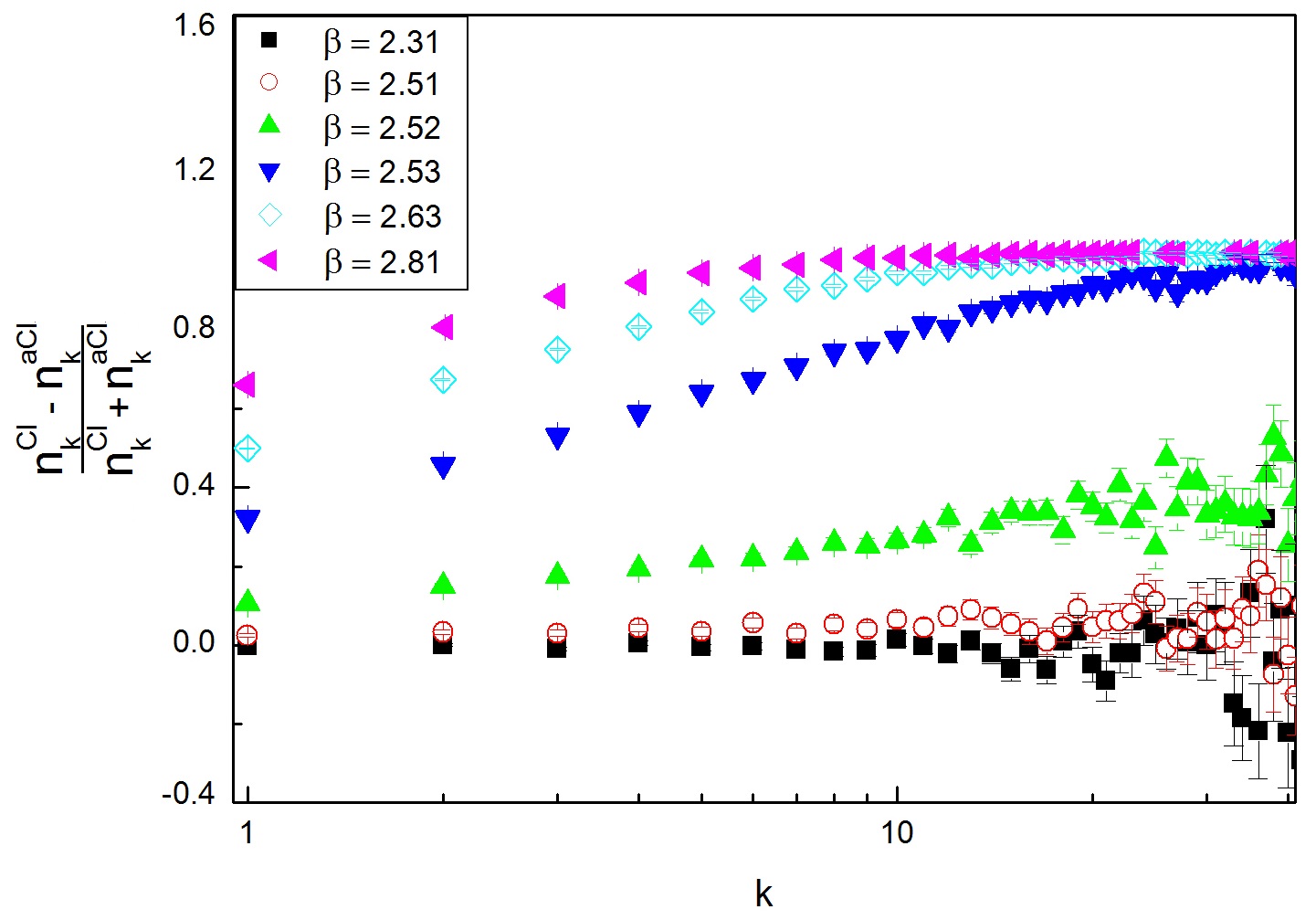}}
\caption{(Color online) The relative asymmetry  
$\frac{n_{cl} (k) - n_{acl} (k) }{n_{acl} (k) + n_{cl} (k)}$ 
of the cluster-anticluster multiplicity as function of the (anti)cluster 
volume $k$ is shown for several values of $\beta$. All results correspond to  
the cut-off $L_{cut}=0.2$.
  }
\label{fig11}
\end{figure}
In the upper vicinity of $\beta_c^\infty$ the inequality (\ref{EqXI}) remains valid, 
while the inequality (\ref{EqXII}) breaks down already for monomers and dimers, 
as one can deduce from Fig. \ref{fig11}. This figure indicates that changing 
$\beta$ from $\beta _c^\infty = 2.5115$ to 2.52, i.e. by about $0.4\%$, one obtains a 
sizable change of the asymmetry of cluster-anticluster multiplicity even for 
small (anti)cluster volumes $k \le 10$. Such an asymmetry can be expressed in the
form of Eq. (\ref{EqIX}) for $q_A = n_{A} (k)$,    
Therefore, Eq. (\ref{EqX}) can be further approximated 
in the vicinity of  $\beta _c^\infty$ as
\begin{eqnarray}\label{EqXIII}
\hspace*{-4.4mm}
{\textstyle \langle L (\vec x)}   \rangle &\simeq  & {\textstyle
 \frac{\widetilde{\sum\limits_{k}} \left[ |L_k^{acl}| + |L_k^{cl}| \right] k \, (n_{acl} (k)-  n_{cl} (k)) + 
 \left[ |L_{max}^{acl}|  +  |L_{max}^{cl}|  \right] (K_{max}^{acl}-K_{max}^{cl})}{2\, N_\sigma^3}}
 \, \quad \quad\\
 \label{EqXIV}
\hspace*{-4.4mm}  &\simeq  &  {\textstyle \frac{ 
 \left[ |L_{max}^{acl}|  +  |L_{max}^{cl}|  \right] (K_{max}^{acl}-K_{max}^{cl})}{2\, N_\sigma^3}}
  \, ,
\end{eqnarray}
where in the last step both sums over all (anti)cluster volumes were neglected 
compared to two contributions of the largest ones. 
This is a good approximation,  since in the vicinity of $\beta_c^\infty$ the 
total volume occupied by the gases of clusters and anticlusters is only a few percents
of the corresponding liquid droplet. 
Therefore, in calculating the mean value of Polyakov loop (\ref{EqX}) we could 
from the very beginning perform a summation only over the largest cluster and 
anticluster. Eq. (\ref{EqXIV}) shows that in the vicinity of $\beta_c^\infty$ 
the volume average of the Polyakov loop $\langle L(\vec x) \rangle$ of SU(2) 
gluodynamics is dominated by the difference between the mean volume of largest 
anticluster and the mean volume of largest cluster. 

Based on the analogy between the volume average 
of the Polyakov loop in SU(2) gluodynamics and the spontaneous magnetization in 
the $Z(2)$ 3-dimensional spin model, we have studied the behavior of two additional 
order parameters in the upper vicinity of $\beta_c^\infty$.
The first of them is the difference 
\begin{eqnarray}\label{EqXV}
\Delta \, {\max{K}_A} (\beta) =  \max{K}_A (\beta) - \max{K}_A (\beta=2.52) \,.
\end{eqnarray}
From Fig. \ref{fig8} one immediately can deduce that  for the largest anticluster 
$\Delta \, {\max{K}_{acl}} (\beta) \ge 0 $ for $\beta \ge 2.52$, while for 
the largest cluster $\Delta \, {\max{K}_{cl}} (\beta) \le 0$ for $\beta \ge 2.52$.  
In order to determine the behavior of the difference (\ref{EqXV})  in the vicinity 
of transition region, we fitted it with two parameters $a_A$ and $b_A$ using the 
formula 
\begin{eqnarray}\label{EqXVI}
\Delta \, {\max{K}_A} (\beta \ge 2.52) =  \pm a_A \cdot(\beta - 2.52)^{b_A}  \,,
\end{eqnarray}
where the sign $+$ ($-$) corresponds to $A = acl$ ($A = cl$). Note that such an 
expansion for the right 
hand side vicinity 
of $\beta=2.52$ was chosen 
because this point, indeed, belongs to the transition region, as one can  
see from Figs. \ref{fig8} and \ref{fig11}, while in our finite system  
the states which correspond to the point $\beta=\beta_c^\infty$ still belong 
to the phase with unbroken symmetry. The results of fit are presented in Table II.
\begin{table}[htp]
\caption{The fit parameters according to Eq. (\ref{EqXVI}).}
\begin{center}
{\footnotesize
\begin{tabular}{|c|c|c|c|c|}
\hline
~ {\rm  cut-off}~  &  ~ {\rm  type}~   &  $a_A$     &    $b_A$   &   $\chi^2 / dof$    \\
\hline
  ~~ $L_{cut} = 0.1 $ ~~ &  ~ {\rm clusters} ~ &~~$3056 \pm 246$~~ &   ~~$0.2964 \pm 0.0284$~~  &  ~~$16.32/4 \simeq 4.08$~~ \\ \hline 
$L_{cut} = 0.1 $   &    ~ {\rm anticlusters} ~ &  $2129 \pm 160$  &   $0.3315 \pm 0.0269$  & ~~$8.94/4 \simeq 2.235$~~ \\ \hline 
$L_{cut} = 0.2 $ &     ~ {\rm clusters} ~ & $4953 \pm 443$  &   $0.3359 \pm 0.0289$ &  $12.3/3 \simeq 4.01$ \\ \hline 
$L_{cut} = 0.2 $&     ~ {\rm anticlusters} ~ & $2462 \pm 87.7$  &    $0.3750 \pm 0.0129$  & $2.068/4 \simeq 0.517$   \\ \hline 
\end{tabular}
}
\end{center}
\label{table2}
\end{table}

{During the fitting we employed the data for 4, 5, 6, 7, 8 and 9 values 
of $\beta$ for each quantity analyzed. Then we have chosen those sets, which provide   
the minimal value of $\chi^2 / dof$. Usually, these were 6 values of $\beta$ beginning 
at $\beta=2.52$, but in one case the lowest value of $\chi^2 / dof$ was achieved for 
five points (see the fourth row from above in Table II). Although in some cases the 
obtained $\chi^2 / dof$ values are somewhat large, the overall quality of the data 
description is good, as one can see from the curves depicted in Fig. \ref{fig8}. 
The obtained large $\chi^2 / dof$ values  can be easily understood, if one takes into 
account  rather small statistical errors and the finite, and not large, size of the 
system under investigation.}  
It is, nevertheless, remarkable that the found  exponents $b_A$ from Table II  are 
very close to the critical exponent $\beta_{\rm Ising} = 0.3265\pm 0.0001$ 
\cite{Beta:3d} of 3-dimensional Ising model  and to the critical exponents 
$\beta_{\rm liquids} = 0.335\pm 0.015$ \cite{Beta:Liquid} of simple liquids, 
although the exponent $b_{acl}$ for the cut-off $L_{cut}=0.2$ is somewhat larger.  

Now we investigate the $\beta$-behavior of the reduced surface tension coefficients 
$\sigma_A(\beta)$ using the following difference
\begin{eqnarray}\label{EqXVII}
\Delta \, {\sigma_A} (\beta \ge 2.52) =  \sigma_A (\beta) - \sigma_A (\beta=2.52) = \pm d_A \cdot(\beta - 2.52)^{B_A} \,,
\end{eqnarray}
which we expanded in the right hand side vicinity of the point $\beta = 2.52$ 
similarly to Eq. (\ref{EqXVI}). In Eq. (\ref{EqXVII}) the sign $+$ ($-$) 
again corresponds to $A = acl$ ($A = cl$). The constants $d_A$ and $B_A$ were found 
similarly to the case of the fitting the largest cluster 
$\beta$-dependence with Eq. (\ref{EqXVI}). With one exception  the lowest 
values of  $\chi^2 / dof$ were achieved for six values of $\beta \in [2.52; 2.677]$.  
However, we found that anticlusters for the cut-off $L_{cut} = 0.2$ reach 
minimal value $\chi^2 / dof$ for  four values of $\beta$ (see the lowest raw 
in Table III). 
The obtained results are given in Table III, which also shows that the quality 
of the fit is  better than for Eq. (\ref{EqXVI}). 
The curves representing Eq.  (\ref{EqXVII}) with the best parameters are shown 
in Fig. \ref{fig6}. 
We can notice that the parameterization 
(\ref{EqXVII}) perfectly describes the reduced surface tension of clusters 
for all values of $\beta \ge 2.677$, i.e. outside the range of fitting. 
This one can see in Fig. \ref{fig6} for the cut-off $L_{cut} = 0.2$. 
The situation for the cut-off $L_{cut} = 0.1$ is absolutely the same.

From Table III one can see that the values of exponents $B_{acl}$ for 
gas of anticlusters are (within error bars) close to 0.5, while the 
exponents $B_{cl}$ for the gas of clusters are 
close to 0.29. In other words, within error bars the exponents $B_{cl}$ 
are close to the critical exponent $\beta_{\rm Ising}$ of 3-dimensional 
Ising model, while $B_{acl}$ are close to the critical exponent of 
mean-field models $\beta_{\rm mf} =0.5$. The large difference between 
values of $B_{acl}$ and $B_{cl}$, however, requires an explanation. 
{Our educated guess is that the exponents $B_{acl}$ for anticlusters 
are affected by the large values of their  chemical potential. }

\begin{table}[htp]
\caption{The fit parameters according to Eq. (\ref{EqXVII}).}
\begin{center}
{\footnotesize
\begin{tabular}{|c|c|c|c|c|}
\hline
~ {\rm  cut-off}~  &  ~ {\rm  type}~   &  $d_A$     &    $B_A$   &   $\chi^2 / dof$    \\
\hline
  ~~ $L_{cut} = 0.1 $ ~~ &  ~ {\rm clusters} ~ &~~$0.485 \pm 0.014$~~ &   ~~$0.2920 \pm 0.0012$~~  &  ~~$1.43 /4 \simeq 0.36$~~ \\ \hline 
$L_{cut} = 0.1 $   &    ~ {\rm anticlusters} ~ &  $2.059 \pm 0.028$  &   $0.4129\pm 0.0077$  & ~~$1.68/4\simeq 0.48 $~~ \\ \hline
$L_{cut} = 0.2 $ &     ~ {\rm clusters} ~ & $0.2796 \pm 0.0118$   &   $0.2891 \pm 0.0016$ &  $1.11/4 \simeq  0.28$ \\ \hline 
$L_{cut} = 0.2 $&     ~ {\rm anticlusters} ~ & $1.344 \pm 0.033$  &    $0.4483 \pm 0.0021$  & $0.66/2 \simeq 0.33$   \\ \hline 
\end{tabular}
}
\end{center}
\label{table3}
\end{table}

From Eqs. (\ref{EqXVI}) and (\ref{EqXVII}) one can obtain two scaling laws 
relating ${\max{K}_A}(\beta)$ and ${\sigma_A}(\beta)$ in the right hand side  vicinity  
 of the point $\beta = 2.52$
\begin{eqnarray}\label{EqXVIII}
\left| \frac{{\max{K}_A} (\beta ) - {\max{K}_A} (2.52) }{a_A }\right|^\frac{1}{b_A} & = & \left| \frac{{\sigma_A} (\beta ) - {\sigma_A} (2.52) }{d_A }\right|^\frac{1}{B_A}   \,.
\end{eqnarray}
These scaling laws allow us to explicitly reexpress the spatial average value 
of the Polyakov loop from  Eq. (\ref{EqXIV}) in terms of  the reduced surface 
tension coefficients ${\sigma_A} (\beta ) $.

\section{Conclusions}

In this work we have studied the phase transformation occurring  in SU(2) 
gluodynamics using the clusters and anticlusters constructed from the Polyakov 
loops. At present the interest devoted to Polyakov loops clusters has been renewed and extended to full QCD with dynamical fermions, and some new aspects have been added 
\cite{Regensburg15} which  make these clusters interesting for physics of 
 heavy ion collisions. In order to investigate the anatomy of deconfining  PT 
in a finite system we have applied the knowhow to describe clusters in the form
of the liquid droplet model which is successfully used in theoretical studies 
of ordinary liquids, nuclear matter and  spin systems. 
However, in contrast to these studies here we have analyzed a new type of objects,
namely,   the clusters formed by Polyakov 
loops of positive and negative  signs. Also an important difference with previous applications of 
the liquid droplet model is that we did not preset the value of Fisher exponent 
$\tau$, but found its value from the requirement of best description of lattice  data generated for all values of $\beta$ studied here. This requirement 
has allowed us to determine the universal value of $\tau \simeq 1.806 \pm0.008$ both for 
clusters and for anticlusters.  This result is in contrast with the famous 
Fisher droplet model \cite{Fisher-67} which assumes $\tau > 2$, but is in line with the predictions of 
two exactly solvable statistical models with a tricritical 
endpoint \cite{Reuter_01,Ivanytskyi}. Our  analysis has showed 
 that the liquid droplet formula works well starting from dimers, only the 
monomers are not satisfactory described by it. 

Our treatment  shows that in the phase of unbroken $Z(2)$ symmetry among the SU(2) 
Polyakov loop clusters  and anticlusters there are two large droplets of almost the same volume. 
According to the present  framework they are considered as 
two different liquids. However, they are not homogeneous inside, but are filled 
by the  smaller clusters which have the opposite sign of the Polyakov loops. 
{Therefore, visually these liquids resemble two pieces of cheese of different sort.}

These internal defects we have interpreted as a gas and applied to its description 
the liquid droplet  model formula. Fit of the size distribution functions for the 
gas of clusters and the gas of anticlusters have allowed us to determine their 
reduced chemical potentials, surface tensions and overall normalization factors. 
As expected, in the symmetric phase the thermodynamic parameters of clusters and 
anticlusters are the same within error bars. 
The situation, however, drastically changes for $\beta > 2.5115$. The largest
droplet (largest anticluster) increases in size and the total number of clusters 
inside it also increases, 
while the next largest droplet, which has an opposite sign of Polyakov loop, 
(largest cluster) shrinks and the number of anticlusters inside it also diminishes. 
Therefore, from the point of view of statistical mechanics it is correct to call 
these processes as condensation and evaporation, respectively.
To our best knowledge this is the first simultaneous analysis of
the collective properties of the two kinds of liquid droplets and their gases 
in SU(2) gluodynamics.

We have demonstrated that 
 the deconfinement PT  in SU(2) gluodynamics  can be easily recognized  by the different behavior of 
chemical potentials, surface tensions and overall normalization factors characterizing 
 the two gases. Thus, the reduced surface tension of the gas of anticlusters 
increases with $\beta$, while the one of clusters gradually decreases and vanishes 
at $\beta > 3.$ A similar behavior is found also  for the corresponding chemical 
potentials. Hence, we conclude that above $\beta =\beta_c^\infty = 2.5115$,  the two gases are 
not in a chemical equilibrium with each other. Our analysis shows that 
all studied thermodynamic quantities characterizing the Polyakov loop clusters 
have bifurcation point at $\beta =\beta_c^\infty$,
i.e. their $\beta$ derivatives on both sides of the point $\beta =\beta_c^\infty$ 
are not equal to each other. 
Here we did not study the thermodynamic limit of the system under investigation and, 
hence,  we cannot determine  an exact value of the (anti)cluster surface 
tension in physical units.
Nevertheless, our approach is able to predict the functional behavior of the 
surface tension in physical units for the gas of clusters at high temperatures.  
It is $\sigma^{phys}_{cl} (T) \simeq T^2 $.  For anticlusters at high temperatures 
we found $\sigma^{phys}_{acl} (T)~  \simeq ~T^4$, but we believe that this function 
may include also  an additional dependence on the reduced chemical potential $\mu_{acl}$.

We have found an  approximate relation between the spatial average of the
Polyakov loop in terms of the difference between the volume of the largest anticluster  
$\max{K_{acl}(\beta)}$ and the volume of the largest cluster $\max{K_{cl}(\beta)}$. 
The investigation of the $\beta$-dependence of the two volumes $\max{K_{A} (\beta)}$ 
in the right hand side vicinity of the point $\beta =2.52$ has allowed us to determine 
the exponents $b_A$, which within the error bars coincide with the critical exponent 
$\beta_{\rm Ising} = 0.3265\pm 0.0001$ \cite{Beta:3d} of 3-dimensional Ising model  
and with  the critical exponents $\beta_{\rm liquids} = 0.335\pm 0.015$ \cite{Beta:Liquid} 
of simple liquids.
Similarly, we determined the critical exponents of the reduced surface tension 
coefficient $\sigma_A (\beta)$ in the right hand side vicinity of the point 
$\beta =2.52$. This has allowed us to establish the scaling laws between 
$\sigma_A (\beta)$  and  $\max{K_{acl}(\beta)}$, which help us to express 
the spatial average of the Polyakov loop  in terms of the reduced surface 
tension coefficient $\sigma_A$
of clusters and anticlusters. Thus, we explicitly show that the reduced surface 
tension coefficients $\sigma_{cl}$ and $\sigma_{acl}$
can be used as the order parameter of this PT.

Also we found an intriguing  conservation law of volume fraction of auxiliary vacuum 
(see Fig. \ref{fig1}). In other words,  for a given cut-off the volume fraction of vacuum 
is independent of $\beta$. This interesting phenomenon, however, requires
further investigation.

In conclusion we want to stress that applying the LDM to the description 
of the (anti)clusters formed by the Polyakov loops gave us many 
interesting and unexpected results.
In particular, the existing exactly solvable models of 
physical clusters should  be further  developed in order to be applicable for  the SU(3) 
gluodynamics.
Obviously,  studies of clustering phenomena 
 in the QCD PT should be continued  within  more realistic models  including
 quark degrees of freedom.

\vspace*{2.2mm}

{\bf Acknowledgements.} Authors thank D. B. Blaschke, O. A. Borisenko, V. Chelnokov, Ch. Gattringer, D. H. Rischke, L. M. Satarov, H. Satz and E. Shuryak for the fruitful discussions and valuable comments. A.I.I., K.A.B., D.R.O., V.V.S., V.K.P. and G.M.Z. acknowledge a partial financial  support of this work 
by  the Program 
 of Fundamental Research of the Department of Physics and Astronomy of  National Academy of Sciences of Ukraine  and  by the National Academy of Sciences of Ukraine Grant of GRID simulations for high energy physics.  I.N.M. acknowledges  a partial support 
 from the Helmholtz International Center for FAIR (Germany) and by the grant NSH-932.2014.2  of the Ministry of Education and Science of the Russian Federation. 
 


\begin{thebibliography}{999}

\bibitem{Svetit:I}
%
L. G. Yaffe and B. Svetitsky, Phys. Rev. D {\bf  26}, (1982) 963.

\bibitem{Svetit:II}
%
L. G. Yaffe, and B. Svetitsky,  Nucl. Phys. B {\bf 210}, (1982) 423.

\bibitem{Polonyi}
%
J. Polonyi and K. Szlachanyi, Phys. Lett. B {\bf 110}, (1982) 395.

\bibitem{Satz:I}
%
S. Fortunato and H. Satz, Phys. Lett. B {\bf 475}, (2000) 311.

\bibitem{Satz:Ib}
%
S. Fortunato {\it et. al.,} Phys. Lett. B {\bf 502}, (2001) 321.


\bibitem{Gatt:I}
%
C. Gattringer, Phys. Lett. B  {\bf 690}, (2010) 179.

\bibitem{Gatt:II}
%
C. Gattringer and A. Schmidt, 
  JHEP {\bf 1101} (2011) 051.


\bibitem{Fisher-67}
%
M. E. Fisher, Physics {\bf 3}, (1967) 255.

\bibitem{Fisher-69}
M. E. Fisher, Rep. Prog. Phys. {\bf 30}, (1969) 615.
	
\bibitem{Stanley:99}
%
H. E. Stanley, 
Rev. Mod. Phys. {\bf 71},  (1999) S358. 
	
\bibitem{RGmethod}
R. Guida and J. Zinn-Justin, J. Phys. Math. Gen. {\bf 31}, (1998)  8103.	

\bibitem{Bondorf}
%
J. P. Bondorf, A. S. Botvina, A. S. Iljinov, I. N. Mishustin, K. Sneppen, Phys. Rep. {\bf 257}, 131 (1995).	


\bibitem{Cluster:1}
%
A. Dillmann and G. E. A. Meier,
J. Chem. Phys. {\bf  94}, (1991) 3872.

\bibitem{Cluster:2}
%
I. J. Ford,
J. Chem. Phys. {\bf 106}, (1997)  9734.
	
\bibitem{Bugaev_00}
%
K. A. Bugaev, M. I. Gorenstein, I. N. Mishustin and W. Greiner, Phys. Rev. {\bf 62} (2000); Phys. Lett. B {\bf 498}, 144 (2001).

\bibitem{Reuter_01}
%
P. T. Reuter, K. A. Bugaev, Phys. Let. B {\bf 517}, 233  (2001); 
Ukr. J. Phys. {\bf 52},   489 (2007). 

\bibitem{GasOfBags:81}
%
J. I. Kapusta,  Phys. Rev.  D {\bf 23}, 2444 (1981).

\bibitem{Bugaev_07}
%
K. A. Bugaev, Phys. Rev. C {\bf 76} (2007) 014903; Phys. Atom. Nucl.  {\bf 71}, 1615 (2008).


\bibitem{Bugaev_09}
%
K. A. Bugaev, V. K. Petrov and G. M. Zinovjev, 
Phys.  Part. Nucl. Lett. {\bf 9},  238 (2012).

\bibitem{Ivanytskyi}
%
A. I. Ivanytskyi, 
Nucl. Phys. A {\bf 880},  12 (2012); 
arXiv:1104.1900 [hep-ph].

\bibitem{Ivanytskyi2}
%
A.~I.~Ivanytskyi, K.~A.~Bugaev, A.~S.~Sorin and G.~M.~Zinovjev,
Phys. Rev. E {\bf 86}, (2012) 061107.

\bibitem{Ivanytskyi3}
%
A. I. Ivanytskyi and K. A. Bugaev,
{Ukr. J. Phys.} {\bf 57}, (2012) 964.

\bibitem{Stock12}
see, for instance, 
F. Becattini, M. Bleicher, T. Kollegger,  M. Mitrovski,  T. Schuster and R. Stock,
Phys.Rev. C85, (2012) 044921  and references therein. 

\bibitem{Shuryak:08}
E. V. Shuryak,
Prog. Part.  Nucl. Phys.  {\bf 62}, (2009) 48.

\bibitem{Bugaev:05}
%
K. A. Bugaev,
Acta. Phys. Polon. B {\bf 36},  (2005)  3083.

\bibitem{Bugaev:07}
%
K. A. Bugaev, 
Phys. Part. Nucl. {\bf 38},  (2007) 447.

\bibitem{Borg}
%
 J.~Borg, I.~N.~Mishustin and J.~P.~Bondorf,
  Phys.\ Lett.\ B {\bf 470} (1999) 13.
  
\bibitem{moretto-03}
%
L. G. Moretto {\it et al.}, Phys. Rev. C {\bf 68}, (2003) 1602, and references therein.


\bibitem{moretto-05}
%
L. G. Moretto {\it et al.}, 
Phys. Rev. Lett. {\bf 94},  (2005) 202701 and references therein.

\bibitem{sagun-14}
%
V. V. Sagun, A. I. Ivanytskyi, K. A. Bugaev and I. N. Mishustin,
Nucl. Phys. A {\bf 924}, (2014) 24.
  
\bibitem{THill:1}
%
T. L. Hill,  {\it Thermodynamics of small  systems}, New York: Dover,  1994.

\bibitem{Moretto:97}
L. G. Moretto {\it et al.},  Phys. Rep.  {\bf 287}, 249 (1997).


\bibitem{DGross:1}
%
D. H. E. Gross, {\it Microcanonical Thermodynamics: Phase Transitions in Finite Systems}, Lecture Notes in Physics,  {\bf 66}, Singapor: World Scientific, 2001. 

\bibitem{Bmodal:Chomaz01}
%
Ph. Chomaz,  F. Gulminelli and  V. Duflot,  Phys. Rev. E.  {\bf  64}  (2001) 046114.


\bibitem{Bmodal:Lopez06}
%
O. Lopez   and  M. F. Rivet,  
Eur. Phys. J.  A. {\bf  30}, (2006) 263   and references therein. 

\bibitem{Fractals}
%
M. I.  Polikarpov, Phys. Usp. {\bf   38}, (1995) 591.

\bibitem{Taylor82}
%
J. R. Taylor, {\it ``An introduction to error analysis'', University Science Book Mill Valley, California (1982).}


\bibitem{Ref_beta_crit}
%
J. Fingberg, U. Heller  and F. Karsch,
  Nucl.\ Phys.\ B {\bf 392}, (1993) 493.


\bibitem{Beta:values}
%
C. Gattringer and C. Lang,
{\it ``Quantum Chromodynamics on the Lattice''}, Springer, Berlin, 2010.

\bibitem{Karsch:2003jg}
  F.~Karsch and E.~Laermann,
  In Hwa, R.C. (ed.) et al.:{\it ``Quark gluon plasma"} (2003) 1; 
  [hep-lat/0305025].

\bibitem{Jaffe:1}
E. Farhi  and R. L. Jaffe,
Phys. Rev. D  {\bf 30}, (1984)  2379.

\bibitem{Svet:1} 
I. Mardor and B. Svetitsky ,
Phys. Rev. D  {\bf 44}, (1991) 878.

\bibitem{Madsen} 
G. Neergaard  and J. Madsen,
Phys. Rev. D  {\bf 62} (2000) 034005.



\bibitem{SurfT:s1}
%
L. F. Palhares and E. S. Fraga, Phys. Rev. D {\bf  82}, (2010) 125018.
 

\bibitem{SurfT:s2}
%
M. B. Pinto, V. Koch  and J. Randrup, 
  Phys.\ Rev.\ C {\bf 86}, (2012) 025203. 

\bibitem{SurfT:m1}
%
D. N. Voskresensky, M. Yasuhira, and T. Tatsumi, Nucl.
Phys. A723, 291 (2003).

\bibitem{SurfT:m2}
%
K. A. Bugaev  and G. M. Zinovjev,
Nucl. Phys. A {\bf 848}, (2010),  443.

\bibitem{SurfT:l1}
%
M. G. Alford, K. Rajagopal, S. Reddy, and F. Wilczek,
Phys. Rev. D {\bf  64},  (2001) 074017.

\bibitem{Gattr_fresh}
%
G. Endrodi, C. Gattringer and H.-P. Schadler,
Phys. Rev. D {\bf  89}, (2014) 054509. 

\bibitem{Greiner07}
I. Zakout, C. Greiner, and J. Schaffner-Bielich, Nucl. Phys. A
781,  (2007) 150.

\bibitem{Greiner08}
I. Zakout and C. Greiner, Phys. Rev. C 78,  (2008) 034916.


\bibitem{Beta:3d}
%
M. Campostrini, A. Pelissetto, P. Rossi and E. Vicari, Phys. Rev. E  {\bf   65},  (2002) 066127.

\bibitem{Beta:Liquid}
%
K. Huang, {\it Statistical Mechanics}, Wiley, New York, 1987.

\bibitem{Regensburg15}
%
 A.~Sch\"afer, G.~Endrodi and J.~Wellnhofer,
  Phys.\ Rev.\ D {\bf 92}, (2015)   014509.




\end{thebibliography}
\end{document}